\documentclass{aa} 

\usepackage{graphicx}
\usepackage{txfonts}
\usepackage{hyperref}
\usepackage{bm}
\usepackage{physics}

\newcommand\bx{{\bm x}}
\newcommand\by{{\bm y}}

\newcommand\bk{{\bm k}}
\newcommand\bl{{\bm l}}
\newcommand\bmm{{\bm m}}

\newcommand\Cref{C_{\text{ref}}}
\newcommand\cref{\Cref}

\newcommand\EE{{\mathbb{E}}}

\DeclareMathOperator{\Var}{Var}

\newcommand\fNL{f_{\rm NL}}
\newcommand\hatfNL{\hat f^b_{\rm NL}}

\newcommand\hatfNLmax{\hat f_{\rm NL}^{\rm filt,max}}

\newcommand\hatbmax{\hat b_{\rm max}}
\newcommand\skewTrue{\gamma_1}
\newcommand\skewMap{\gamma}
\newcommand\skewData{\bar{\skewMap}}
\newcommand\skewFiltMax{\skewMap^{{\rm filt, max}}}
\newcommand\skewB{\hat{\skewMap}^B}
\newcommand\hatmudata{\skewData}
\newcommand\hatmub{\skewB}
\newcommand\hatmufiltmax{\skewFiltMax}

\newcommand\percent{\%}

\newcommand{\XG}{X_{\rm G}}
\newcommand{\XNL}{X_{\rm NL}}

\usepackage{xcolor}

\newcommand{\VB}[1]{{#1}}

\begin{document} 

\title{{Characterizing the spatial pattern of solar supergranulation using the bispectrum}}

\author{
        Vincent G. A. Böning\inst{1}
        \and
        Aaron C. Birch\inst{1}
        \and
        Laurent Gizon\inst{1,2,3}
        \and
        Thomas L. Duvall, Jr.\inst{1}
        \and
        Jesper Schou\inst{1}
}

\institute{
        {Max-Planck-Institut f\"ur Sonnensystemforschung, Justus-von-Liebig-Weg 3, 37077 G\"ottingen, Germany}\\
        \email{boening@mps.mpg.de}
        \and
        {Georg-August-Universit\"at, Institut f\"ur Astrophysik, Friedrich-Hund-Platz 1, 37077 G\"ottingen, Germany}
        \and
        {Center for Space Science, NYUAD Institute, New York University Abu Dhabi, PO Box 129188, Abu Dhabi, UAE}
        }

\date{Received 17 December 2019; Accepted 5 February 2020}


\abstract
{The spatial power spectrum of supergranulation does not fully characterize the underlying physics of turbulent convection. For example, it does not describe the \VB{non-Gaussianity} in the horizontal flow divergence.}
{Our aim is to statistically characterize the spatial pattern of solar supergranulation beyond the power spectrum. {The next-order statistic is the bispectrum. It measures correlations of three Fourier components and is related to the nonlinearities in the underlying physics. It also characterizes how a skewness in the dataset is generated by the coupling of three Fourier components.}
}
{We estimated the bispectrum of supergranular horizontal surface divergence maps that were obtained using local correlation tracking (LCT) and time-distance helioseismology (TD) from one year of data from the Helioseismic and Magnetic Imager on-board the Solar Dynamics Observatory starting in May 2010.}
{We find significantly nonzero and consistent estimates for the bispectrum using LCT and TD. The strongest nonlinearity is present when the three coupling wave vectors are at the supergranular scale. These are the same wave vectors that are present in regular hexagons, which have been used in analytical studies of solar convection. At these Fourier components, the bispectrum is positive, consistent with the positive skewness in the data and consistent with supergranules preferentially \VB{consisting of} outflows surrounded by a network of inflows. We use the bispectral estimates to generate synthetic divergence maps that are very similar to the data. This is done by a model that consists of a Gaussian term and a weaker quadratic nonlinear component. Using this method, we estimate the fraction of the variance in the divergence maps from the nonlinear component to be of the order of 4-6\%.}
{We propose that bispectral analysis is useful for understanding the dynamics of solar turbulent convection, for example for comparing observations and numerical models of supergranular flows. This analysis may also be useful to generate synthetic flow fields.}


\maketitle


%


\section{Introduction}
\label{secIntro}

Supergranular motions are characterized by horizontal velocities of about $300 \,\rm{m}\,\rm{s}^{-1}$ and are coherent features with a length scale of about $30\,\rm{Mm}$ or harmonic degrees of about $l \sim 120$ in the power spectrum \citep[e.g.,][and references therein]{Rincon2018}. Close to the surface, supergranules are observed as regions of flow with horizontal divergence (associated {with} upflows) surrounded by a network of convergent flows \citep[associated with downflows, e.g.,][]{Duvall2000}.

The origin of supergranulation as a dominant scale of convection remains unclear \cite[see][for a review]{Rincon2018}. 
Supergranulation as a particular length-scale might emerge because of thermal instabilities in the ionization zones \citep{Simon1964}, as the largest buoyantly driven scale of convection in the Sun \citep{Cossette2016}, or the largest convective scale not significantly constrained by rotation
\citep{Featherstone2016}. 
Also unexplained is the evolution of the supergranulation pattern, which displays a wave-like behavior \citep{Gizon2003,Schou2003} -- although this is not the topic of this paper.

Comparisons between theory and observations of {supergranular-scale} flow fields are {usually} based on the power spectrum of horizontal surface flows {\citep[e.g.,][]{Rincon2017}}. However, it is clear that the power spectrum does not contain much information about the spatial pattern of the flow field, apart from the scale-dependence of the amplitude. For example, the power spectrum does not contain any information about the skewness of the underlying field \citep[{see e.g.,} Figure~5 in][]{Nordlund1997}. Divergence maps of supergranulation have a positive skewness \citep{Duvall2000} which is due to the asymmetric distribution of upflows and downflows.

Such an asymmetry between up- and downflows is present in idealized hexagonal convection cells that have been used in analytical studies of convection and supergranulation in particular \citep[e.g.,][]{Chandrasekhar1961,Gough1975,Busse2007}. These hexagonal cells are characterized by a central upflow region surrounded by a network of downflows \citep{Chandrasekhar1961,Toomre1977,Busse2003}, similar to supergranulation. The opposite case of strong downflows surrounded by a weaker network of upflows was also considered \citep{Busse2004,Busse2007}. 
For example, theoretical attempts to explain the phase velocity of supergranulation and its contribution to maintaining the near-surface shear layer \citep{Busse2004,Busse2007} are based on decomposing the solution to the Navier-Stokes equations in terms of planforms that are constructed from such patterns in the horizontal direction.

The main aim of our work is to provide a statistical characterization of the spatial pattern of supergranulation beyond the power spectrum. 
In other words, we want to find out whether convective patterns, such as hexagons, can be directly related to the supergranulation data. 
Additionally, our longer-term aim is to construct synthetic supergranular flow fields. The availability of synthetic flow fields could be useful in helioseismic methods for studying subsurface convection \citep[e.g.,][]{Hanasoge2012,Duvall2014}.

Information on the skewness is contained in the bispectrum, which is a next-order generalization of the power spectrum \citep[e.g.,][Section 11.5.1]{Priestley1981}. Just as the power spectrum describes the variance as a function of scale, the bispectrum describes the skewness as a function of two coupling scales \citep[e.g.,][]{Watkinson2017}. Measuring the bispectrum of supergranular divergence maps would therefore be useful to characterize the skewness as a function of scale. Furthermore, the bispectrum is directly related to hexagonal patterns because a hexagon has a specific signature in the bispectrum. The bispectrum characterizes the coupling of two wave vectors to a third wave vector, which appears similarly in the nonlinearities in the Navier-Stokes equations. Such nonlinearities are present in the modal equations of convection studied using horizontal patterns \citep[e.g.,][]{Chandrasekhar1961,Gough1975,Busse2007}.

To clarify the results obtained in this paper, it is necessary to clarify what we mean by linear and nonlinear. By linear, we mean the hypothesis that the time-evolution of supergranular divergence maps was given by a linear partial differential equation (PDE) and a stochastic forcing term with independent Gaussian Fourier components. In this case, the dynamics at the Fourier components of the solution would also be independent. By nonlinear, we mean the hypothesis that the dynamics of supergranular divergence maps was described by a PDE that in addition includes a nonlinear term. In this case, the Fourier components of the solution would not be independent but would show some dependence that might be measurable. Supergranular dynamics are strongly nonlinear {according to} \citet{Rincon2017} and divergence maps have a large nonzero skewness \citep{Duvall2000}. This should be reflected in the bispectrum.

Introductions to bispectral analysis are given by \citet{Brillinger1965}, \citet{Kim1979}, \citet[][Section 11.5.1]{Priestley1981}, \citet{Nikias1987}, and \cite{Chandran1990}. Bispectral analysis has been used in a number of research domains. For example, it was used to measure nonlinear wave--wave coupling \citep[e.g.,][]{Kim1980} of ocean and surface water waves \citep[e.g.,][]{Elgar1989}, to study \VB{non-Gaussianity} in the cosmic microwave background \citep[e.g.,][]{Komatsu2001,Ade2016}, and in laboratory plasma physics \citep[e.g.,][]{Itoh2017}.
The bispectrum is related to the spectral transfer of kinetic energy in turbulent flows \citep[e.g.,][]{Yeh1973}. In the context of stellar convective dynamos, measurements of the spectral energy transfer in magnetohydrodynamical simulations have been taken for example by \cite{Brandenburg2001}, \cite{Moll2011}, and \cite{Strugarek2013}.

In this work, we apply bispectral analysis to near-surface horizontal divergence maps of supergranulation obtained by \cite{Langfellner2018} using local correlation tracking \citep[LCT,][]{November1988} and time-distance helioseismology \citep[TD,][]{Duvall1993} and data from the Helioseismic and Magnatic Imager \citep[HMI,][]{Scherrer2012,Schou2012} onboard the Solar Dynamics Observatory \citep[SDO,][]{Pesnell2012}.

\section{Divergence maps of supergranular flows}
\label{secDataIntro}

We use maps of the horizontal near-surface divergence that were obtained by \cite{Langfellner2018} using Dopplergrams and intensity images from HMI/SDO. The data cover 365 days from May 2010 through April 2011.

Regions of size $\sim 180 \times 180 \,\rm{Mm}^2$, which are centered at the equator, were tracked for eleven days from about $70\degr$ east to $70\degr$ west at the surface rotation rate \citep{Snodgrass1984}. Dopplergrams and intensity maps were remapped and each eleven-day time-series was subdivided into 33 eight-hour segments. The seventeenth segment crosses the central meridian.

Using the intensity images and LCT of granules \citep{November1988}, \cite{Langfellner2018} obtained estimates of the two horizontal flow components for the same spatial and temporal coverage. This results in maps with $97\times97$ pixels, each of size $(5\times 0.348\,\rm{Mm})^2$. These maps span a slightly smaller region of size $\sim 170 \times 170 \,\rm{Mm}^2$. \VB{From these maps, w}e computed the horizontal divergence by taking derivatives using Savitzky-Golay filters \citep{Savitzky1964}. In the Savitzky-Golay filters, we used a different filter width (3 instead of 15 pixels) and polynomial order (2 instead of 3) compared to \cite{Langfellner2018} to keep the smoothing as small as possible.

For each segment, Dopplergrams were filtered in the Fourier domain using an $f$-mode ridge filter \citep{Langfellner2015} which includes power from wave numbers $300 < kR_\sun < 2600$ with a peak at $kR_\sun \approx 1100$. {Here, $R_\sun$ is the solar radius and $k$ is the horizontal wave number. Next,} $f$-mode point-to-annulus \citep{Duvall1996} travel-time maps \citep{GB2004} were obtained from the filtered Dopplergrams in the inward minus outward sense. This resulted in maps with $512\times512$ pixels, with each pixel being $(0.348\,\rm{Mm})^2$ in size. These travel-time maps are known to be a proxy of near-surface horizontal divergence of the fluid motion \citep[e.g.,][]{Langfellner2015}. To remove the smallest-scale noise from the data, we low-pass filter this dataset in the Fourier domain and keep Fourier components with $k R_\sun < 1200$.

In addition, we obtained a smoothed version of the (not low-pass filtered) TD maps in order to better compare to earlier results. We used a two-dimensional Gaussian window with full width at half maximum of 5.22 Mm (15 pixels), which is similar to the smoothing done by \cite{Duvall2000}.

\begin{figure*}[ht]
        \centering

        \includegraphics[width=\linewidth]{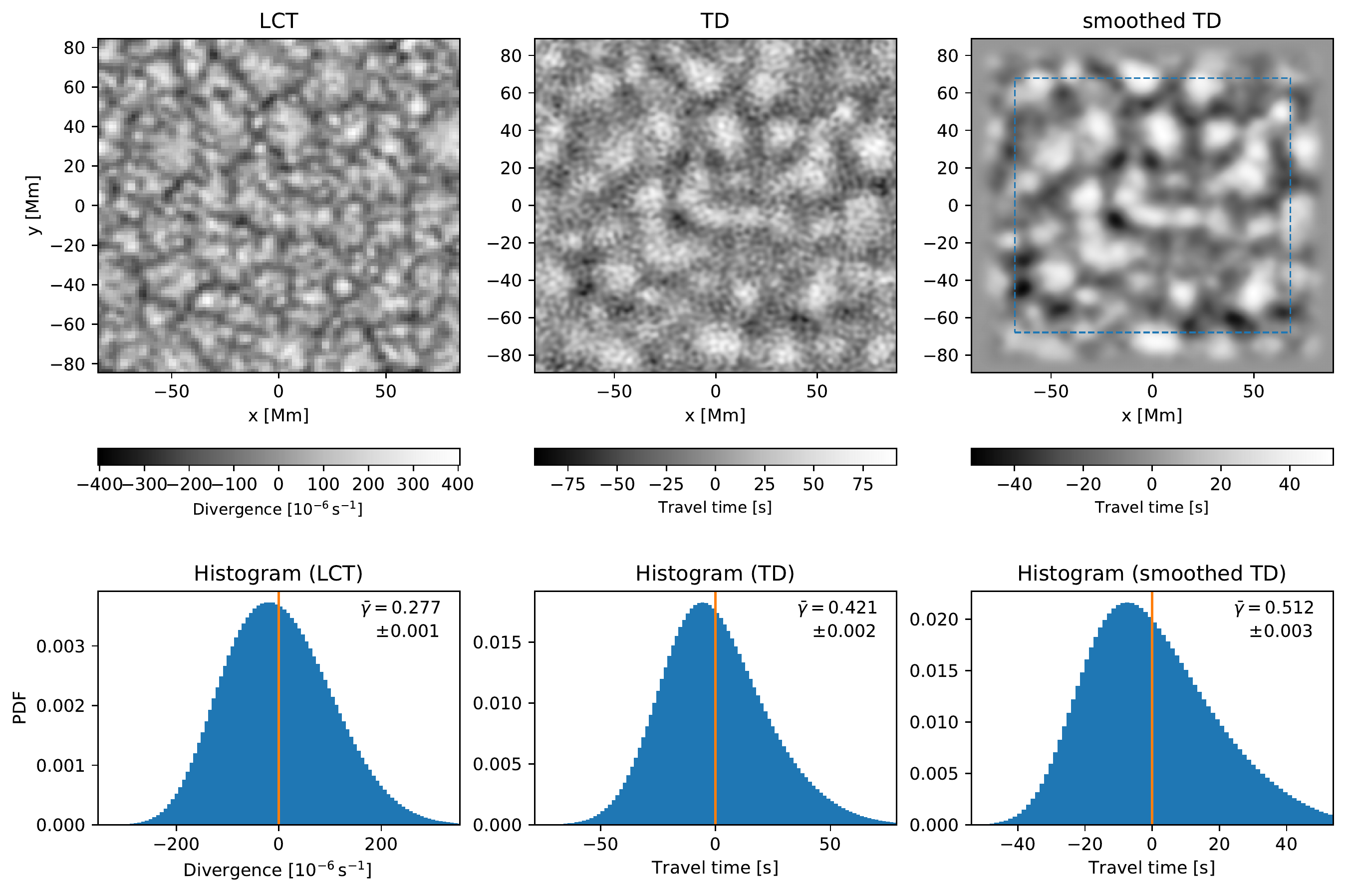}

        \caption{Example divergence maps (top) and corresponding histograms (bottom) for LCT (left), TD (middle), and smoothed TD data (right). The smoothed TD data was apodized with a raised cosine and the histogram was computed for the region inside the dashed rectangle. The histogram and skewness $\skewData$ were computed from the entire dataset.}
        \label{figDataExamples}
\end{figure*}

\VB{For all data sets discussed here, the horizontal mean of each map has been removed.} Figure~\ref{figDataExamples} shows example LCT and TD maps. The most prominent features in these maps are outflows due to supergranules, surrounded by inflows. Figure~\ref{figDataExamples} also shows histograms of the maps. Both LCT and TD histograms are asymmetric about zero and have a nonzero skewness, which is well known \citep[e.g.,][]{Duvall2000}. The skewness of the $i$th map $X^{{(i)}}(\bx)$ is
\begin{align}
\skewMap^{{(i)}} &= \frac{\langle X^{{(i)}}(\bx)^3 \rangle}{\sigma_{X^{{(i)}}}^3},
\end{align}
where the variance is
\begin{align}
\sigma_{X^{{(i)}}}^2 = \langle X^{{(i)}}(\bx)^2 \rangle= h_k^4 \sum_{\bk} |X^{{(i)}}(\bk)|^2.
\end{align}
Here, $\bx=(x,y)$ denotes the horizontal location, $\langle \cdot \rangle$ is a horizontal average, $\langle X^{{(i)}}(\bx)\rangle = 0$ by construction, $X^{{(i)}}(\bk)$ denotes the horizontal Fourier transform of $X^{{(i)}}(\bx)$ -- see Appendix~\ref{appFourier} --, and $h_k$ is the resolution in Fourier space. We use the same symbol for a function and its Fourier transform. We use $\skewData$ for the (mean) skewness computed from the entire dataset.

The skewness depends slightly on disk position. In the following, we therefore only use the 11 locations that are closest to disk center for each segment, ranging from about $23\degr$ east to $23\degr$ west, where the skewness is approximately constant. For the entire datasets, this results in a skewness of $\hatmudata=0.277\pm0.001$ for the LCT maps, $\hatmudata=0.421\pm0.002$ for the TD maps, and $\hatmudata=0.512\pm0.003$ for the smoothed TD maps. \citet{Duvall2000} reported a skewness of 0.42 for TD maps. As we see in Section~\ref{secresults}, different filters applied in the spatial domain may result in different values for the skewness.

The fact that the divergence maps have a nonzero skewness suggests that different Fourier components of the divergence field cannot be independent. If they were independent, the data would have zero skewness, just like a random wave field. As the statistical properties of the supergranular divergence field are translation invariant, it is clear that two Fourier components are uncorrelated. As we see in Section~\ref{secBispectrumIntro}, the nonzero skewness of the data can be related to nonzero correlations between three Fourier components, which leads to the definition of the bispectrum. As such correlations between three Fourier components are also present in simple patterns of convection, we now look at these patterns before introducing the bispectrum.

\section{A brief account of convection patterns}
\label{secPatternsIntro}

We briefly discuss a few patterns used in analytical studies of convection. \VB{Related to the specific case of solar supergranulation}, \citet{Busse2003,Busse2004,Busse2007} studied hexagonal convection cells. This work is based on earlier and more general studies of Boussinesq convection that used convection patterns such as rolls, squares, and hexagons \citep[e.g.,][]{Palm1960,Chandrasekhar1961,Toomre1977} and the decomposition of convective motions into planforms, which are made up from such patterns in the horizontal direction \citep[e.g.,][]{Chandrasekhar1961,Gough1975}.  
Due to the continuity equation and under the assumption of a steady flow, the vertical flows and the horizontal divergence show the same pattern close to the surface.

\begin{figure*}
        \centering
        \includegraphics[width=\linewidth]{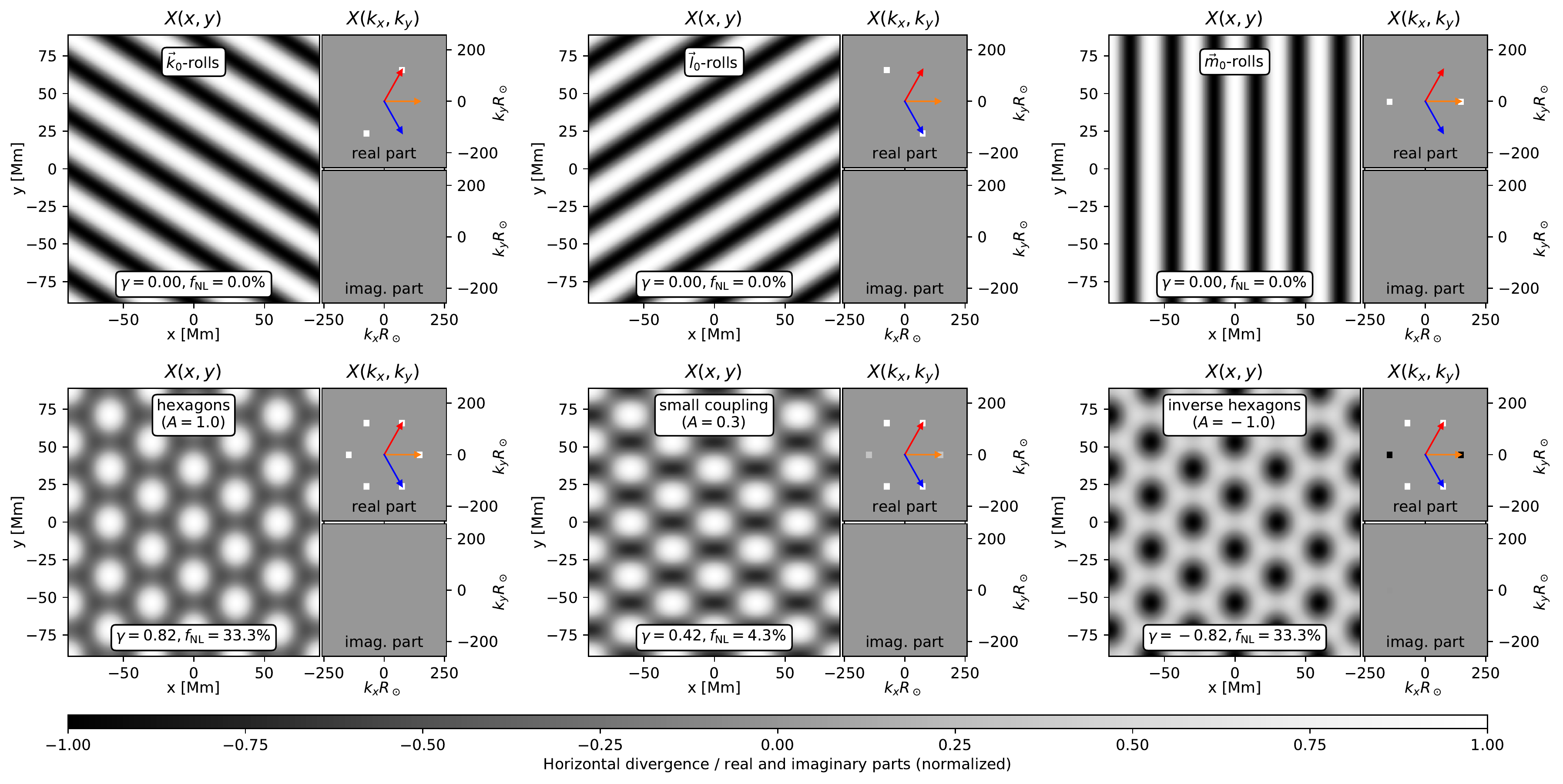}
        \caption{Example patterns $X(\bx)$ and their Fourier transform $X(\bk)$. The wave vectors in the right small panels are: $\bk_0$ (red), $\bl_0$ ({blue}), and $\bmm_0$ (orange). {The skewness $\skewMap$ and fractional power $\fNL$ of the nonlinearly coupled component are given in each panel.} Equation{s~\eqref{eqRolls}} and~\eqref{eqPatterns} give definition{s} of the patterns.}
        \label{figPatterns}
\end{figure*}

We now look at a few possible patterns in detail. The top row in Figure~\ref{figPatterns} shows simple convective rolls. Such a roll is given by a pattern $X(\bx)$ for which
\begin{align}
X(\bx) = h_k^2 \left( e^{i \bk_0 \cdot \bx} + c.c. \right), \label{eqRolls}
\end{align}
 where $\bk_0=(k_{x,0},k_{y,0})$ is a chosen wave vector and we use the notation $+ c.c.$ to indicate that the previous terms are repeated with a complex conjugate. The top row in Figure~\ref{figPatterns} shows three examples of rolls for three different wave vectors $\bk_0,\bl_0,\bmm_0$, which are kept fixed for the entire section and are shown in the small subpanels of Figure~\ref{figPatterns} as colored arrows. These wave vectors form a triad, that is $\bk_0+\bl_0=\bmm_0$. The wave vectors chosen here have the same length and are at an angle of 60 degrees (up to a small deviation caused by the discretization).

We now consider different patterns of the form
\begin{align}
X(\bx) = h_k^2 \left(  e^{i \bk_0 \cdot \bx} + e^{i \bl_0 \cdot \bx}  + A e^{i \bmm_0 \cdot \bx} + c.c. \right), \label{eqPatterns}
\end{align}
where $A$ is the (possibly complex) amplitude of the third Fourier component. This pattern can be seen as two independent rolls ($\bk_0$ and $\bl_0$) plus a nonlinear component ($\bmm_0$). This nonlinear component is given by a quadratic coupling \citep{Kim1979},
\begin{align}
X(\bmm_0)= {A} \, X(\bk_0) X(\bl_0) \label{eqXNLpatterns},
\end{align}
where $A$ is a coupling coefficient. If one thinks of $X(\bmm_0)$ as being due to such a coupling, the nonlinear fraction of the variance is 
\begin{align}
\fNL &= \frac{|X(\bmm_0)|^2}{|X(\bk_0)|^2+|X(\bl_0)|^2+|X(\bmm_0)|^2} = \frac{|{A}|^2}{2 + |{A}|^2},
\end{align}
which is given in the panels in the bottom row in Figure~\ref{figPatterns}.

A hexagonal pattern is given by $A=1$; see bottom left panel in Figure~\ref{figPatterns}. Equation~\eqref{eqPatterns} with $A=1$ is equivalent to the horizontal dependence in Equation (249) in Section 16c of \citet[][]{Chandrasekhar1961}. Hexagonal patterns have the advantage that they can be used to model a convective pattern that is characterized by a central upflow region surrounded by a network of downflows \citep{Chandrasekhar1961,Toomre1977,Busse2003}, similar to supergranulation. 
In addition, hexagonal planforms are not symmetric with respect to a change in sign and they produce a nonzero skewness in the horizontal divergence, similar to the supergranular divergence maps (see Section~\ref{secDataIntro})\VB{;} see the skewness of each pattern given in Figure~\ref{figPatterns}.

The case $A=-1$ corresponds to a pattern of inverted hexagons (see right panel in the bottom row of Figure~\ref{figPatterns}). This corresponds to downflows surrounded by a network of upflows as considered by \citet{Busse2004,Busse2007}.

The case $A=0$ corresponds to rectangles, and we can vary $A$ continuously between zero and one to obtain patterns with different coupling amplitudes, intermediate between rectangles and hexagons; see the example labeled ``small coupling'' ($A=0.3$) in Figure~\ref{figPatterns}.

In addition to these patterns, it is possible to tessellate space by a pattern generated from any combination of three wave vectors with $\bk_0+\bl_0=\bmm_0$, each with a separate, possibly complex Fourier amplitude. For a given pattern, the phases of $X(\bk_0)$ and $X(\bl_0)$ may be jointly put to zero by a translation in real space. Statistically, we do not expect any preferred direction for supergranular divergence maps near the equator, and it is therefore reasonable to assume $|X(\bk_0)|=|X(\bl_0)|$ for illustrative purposes. \VB{The Fourier amplitudes of $X$} may {then} be normalized to one as done above.

For the patterns considered here, one pattern fills the whole domain. On the Sun, however, every supergranule has its individual shape. It is therefore possible that a coupling is present not only at the Fourier components with equal length ($|\bk_0|=|\bl_0|=|\bmm_0|$, as for regular hexagons), but potentially for any coupling triad of Fourier components. As a consequence, the signature in Fourier space is a more complex combination of the previous characteristics at several Fourier components. This motivates the definition of the bispectrum.

\section{Bispectral analysis in two dimensions}
\label{secBispectrumIntro}

Here we introduce and summarize the definition of the bispectrum and related quantities for two-dimensional random processes. This part is mainly based on Section 11.5.1 in \cite{Priestley1981}, \cite{Kim1979}, and \cite{Chandran1990}. Later in this section, we give a few examples and illustrate the definitions with the help of the patterns introduced in the previous section. In addition, we introduce a method for generating data with a given bispectrum.

\subsection{Definition of the bispectrum}
\label{secBispecDef}

Here we consider a random field $X(\bx)$ that depends on a two-dimensional discretely sampled spacial variable $\bx$. We assume that all statistical properties of $X(\bx)$ are translation invariant. This is the generalization of a stationary process to two (spatial) dimensions. Since $X(\bx)$ is a random variable, it induces the definition of an expectation value, which we denote by $\EE[\cdot]$. The spatial Fourier transform of $X(\bx)$ is again denoted by $X(\bk)$. By $X^{(i)}(\bx)$, we denote an individual independent realization of $X(\bx)$, that is, an individual map.

The bispectrum is the next higher-order statistic beyond the power spectrum, and is defined as
\begin{align}
B(\bk,\bl,\bmm=\bk+\bl) &= h_k^2 \, \EE[X(\bk) X(\bl) X^*(\bmm) ]. \label{eqBispecDiscrete}
\end{align}
The three-point correlation function is the Fourier transform of the bispectrum, just as the two-point (auto-)correlation function is the Fourier transform of the power spectrum:
\begin{align}
\EE[X(\bx_0 + \bx) X(\bx_0 + \by)X(\bx_0)] 
&= h_k^4 \sum_{\bk,\bl} B(\bk,\bl,\bk+\bl) 
\,e^{i\bk \cdot \bx + i\bl \cdot \by}, \label{eqE3Bispec}
\end{align}
where the expectation value on the left-hand side of Equation~\eqref{eqE3Bispec} does not depend on $\bx_0$ due to the assumed translation invariance of $X(\bx)$.

Similarly, the skewness of the random field $X(\bx)$, $\skewTrue$, which is the third-order moment normalized by the variance, is the sum of the variance-normalized bispectrum,
\begin{align}
\skewTrue &= \frac{\EE[X(\bx_0)^3]}{\sigma^3} = \frac{h_k^4}{\sigma^3} \sum_{\bk,\bl}  B(\bk,\bl,\bk+\bl) \, , \label{eqSkewnBispec}\\
\sigma^2 &= \EE[X(\bx_0)^2] = h_k^2 \sum_\bk P(\bk), \label{eqSigmaDef}
\end{align}
where the power spectral density $P$ is defined by
\begin{align}
P(\bk) &= h_k^2 \, \EE[|X(\bk)|^2]. \label{eqPowerDef}
\end{align}
Table~\ref{tabFormulas} provides a summary of key terms defined in this paper for later reference.

In the following, we assume that there is no preferred horizontal direction in the data. In this case, the bispectrum of the divergence field is real. The expectation value $\EE[X(\bk) X(\bl) X^*(\bmm) ]$ is zero if $\bmm\neq\bk+\bl$ because of the assumed translation invariance. Therefore, the triadic relation $\bmm=\bk+\bl$ is implicitly understood in the following and the third variable is omitted in bispectrum-related quantities.

\subsection{Estimating the bispectrum}
\label{secEstB}

In principle, the bispectrum can be estimated using Equation~\eqref{eqBispecDiscrete} by simply averaging over {independent} realizations $X^{(i)}(\bk)$ of the data \citep[e.g.,][]{Kim1979}. See Appendix~\ref{appBispecEst} for a formal definition of the estimator $\hat B$, which is obtained this way. 
We take this approach although it may not be optimal; \citet{Nikias1987} discuss other possible approaches.

In order to assess the significance of the estimated bispectrum, it is necessary to consider the bispectral coherence (also known as the bicoherence) $b$, which is a normalized bispectrum \citep[e.g.,][]{Kim1979},
\begin{align}
b(\bk,\bl) 
&= \frac{\EE[X(\bk) X(\bl) X^*(\bmm) ]}{\sqrt{\EE[|X(\bk)X(\bl)|^2] \EE[|X(\bmm)|^2]}}. \label{eqbDef}
\end{align}
The definition of $b$ is independent of the convention used to define the Fourier transform because it is normalized and because the bispectrum is real in our case. We see later that the squared bicoherence $b^2$ quantifies the fraction of power which is due to quadratic nonlinear coupling at $\bmm=\bk+\bl$. For such a coupling to be present in the data and a bispectral measurement to be significant, it is therefore necessary that the bicoherence be significantly different from zero. The variance of an estimator $\hat b$ of the bicoherence that is built by simply averaging {over independent} realizations for estimating expectation values in Equation~\eqref{eqbDef} -- see Appendix~\ref{appBispecEst} --, is bounded by \citep{Kim1979}
\begin{align}
\Var[\hat b(\bk,\bl)] &\approx \frac{1}{N_B} (1-b(\bk,\bl)^2) \leq \frac{1}{N_B},
\end{align}
where $N_B$ is the number of independent data realizations used in the estimate. The statistical $1\sigma$ error in the bicoherence estimate is therefore $\sigma_b = 1/\sqrt{N_B}$. Additionally, we average the results azimuthally -- see Appendix~\ref{appAzimAvg} -- which results in a typical $\sigma_b=0.0035$ at $|\bmm| R_\sun=155.5$.

\subsection{A model with a weak quadratic nonlinearity}
\label{secModel}
\label{secgenerate}

We here present a model that helps us to interpret the measured bispectrum signal and to generate synthetic data that have approximatively a given bispectrum. There are several ways to generate data with a given power spectrum and bispectrum; see for example \cite{Contaldi2001} and \cite{Wagner2010}. Such a method is not unique \citep[e.g.,][]{Wagner2010}. Following \citet{Wagner2010}, we assume that the data can be represented as
\begin{align}
X &= X_{\rm G} + X_{\rm NL} \label{eqXgenerate},
\end{align}
where $\XG$ is a Gaussian field and $\XNL$ a small nonlinear component. In the Fourier domain, the Gaussian field follows a complex Gaussian distribution,
\begin{align}
X_{\rm G}(\bk) &= \mathcal{N}_{\bk} \sqrt{\frac{1}{h_k^2} P(\bk)}, \label{eqXGgenerate}
\end{align}
where $\mathcal{N}_{\bk}$ are Gaussian random variables with independent real and imaginary parts, zero mean and unit variance. We set $\mathcal{N}_{-\bk} = \mathcal{N}_{\bk}^*$ to ensure that $X_{\rm G}$ is real, but otherwise the $\mathcal{N}_{\bk}$ are independent for different $\bk$.

We assume that the nonlinear component is obtained from the Gaussian one by an arbitrary quadratic nonlinearity,
\begin{align}
X_{\rm NL}(\bmm) &= \sum_{\bk} A_{\bk, \bmm -\bk} X_{\rm G}(\bk) X_{\rm G}(\bmm -\bk) \label{eqXNLgenerate},
\end{align}
where $A_{\bk,\bl}$ denote coupling coefficients of modes $\bk$ and $\bl=\bmm-\bk$. To guarantee that $\XNL$ is real, the coupling coefficients have to fulfill $A_{-\bk,-\bl} = A_{\bk,\bl}^*$.

Synthetic data with a given bispectrum $B_0(\bk,\bl)$ can then be generated by setting
\begin{align}
A_{\bk,\bl} &= \frac{1}{6} \frac{h_k^2\, B_0(\bk,\bl) }{ P(\bk)P(\bl)}. \label{eqAkl}
\end{align}
Appendix~\ref{appModel} outlines further details of this model.

\subsection{The bispectrum of hexagons}

For the patterns introduced in Section~\ref{secPatternsIntro}, the bispectrum of the patterns is given by $B(\bk_0,\bl_0) = h_k^2 A$ and the coupling coefficient as defined above is given by $A_{\bk_0,\bl_0} = A/6$. The factor of one-sixth appears because in the model introduced in this section it is assumed that a coupling occurs at several Fourier components. In the case of only one triad of coupling Fourier components as for the patterns from Section~\ref{secPatternsIntro}, one has $A_{\bk_0,\bl_0} = A$.

The bicoherence of the patterns with nonzero coupling coefficient $A$ is $|b(\bk_0,\bl_0)| = 1$. This is because the signal at Fourier component $\bmm_0$ is entirely due to the nonlinear coupling, and there is no noise or any other independent signal at this component.

\subsection{The impact of noise on bispectral estimates}
\label{secNoise}

If noise is present in the data, this reduces the bispectral estimates. If we assume for example that
\begin{align}
X &= X_{\rm G} + X_{\rm NL} + \epsilon \label{eqXnoise},
\end{align}
where $\epsilon(\bx)$ is the noise, which we assume to have independent Gaussian Fourier components and to be independent of $X_{\rm G}$ and $X_{\rm NL}$. If the noise is simply added to the data, the bispectrum (as an expectation value) does not change; see Eqs.~\eqref{eqBispecDiscrete} and~\eqref{eqBispecAkl}. However, the estimated coupling coefficient $A_{\bk,\bl}$ and the bicoherence $b$  are reduced in amplitude because of the increase in power from the noise; see Eqs.~\eqref{eqbDef} and~\eqref{eqAkl}. It is therefore not possible to discriminate between a dataset with small noise and a small coupling coefficient, and another one with larger-amplitude noise and a larger coupling coefficient.

In other words, for a given power spectrum, which is due to the signal plus noise, stronger noise results in a smaller bispectrum. For the case of possible hexagonal convection cells, it might therefore be that their signal in the bispectrum is reduced by noise.

\section{Results}
\label{secresults}

In this section, we show results for data averaged over the azimuthal direction of $\bmm$. We do not find any direction-dependence of the bispectrum.

\subsection{The bispectrum of supergranulation}
\label{secresultsbispec}

\begin{figure*}
        \centering
        \includegraphics[width=0.32\linewidth]{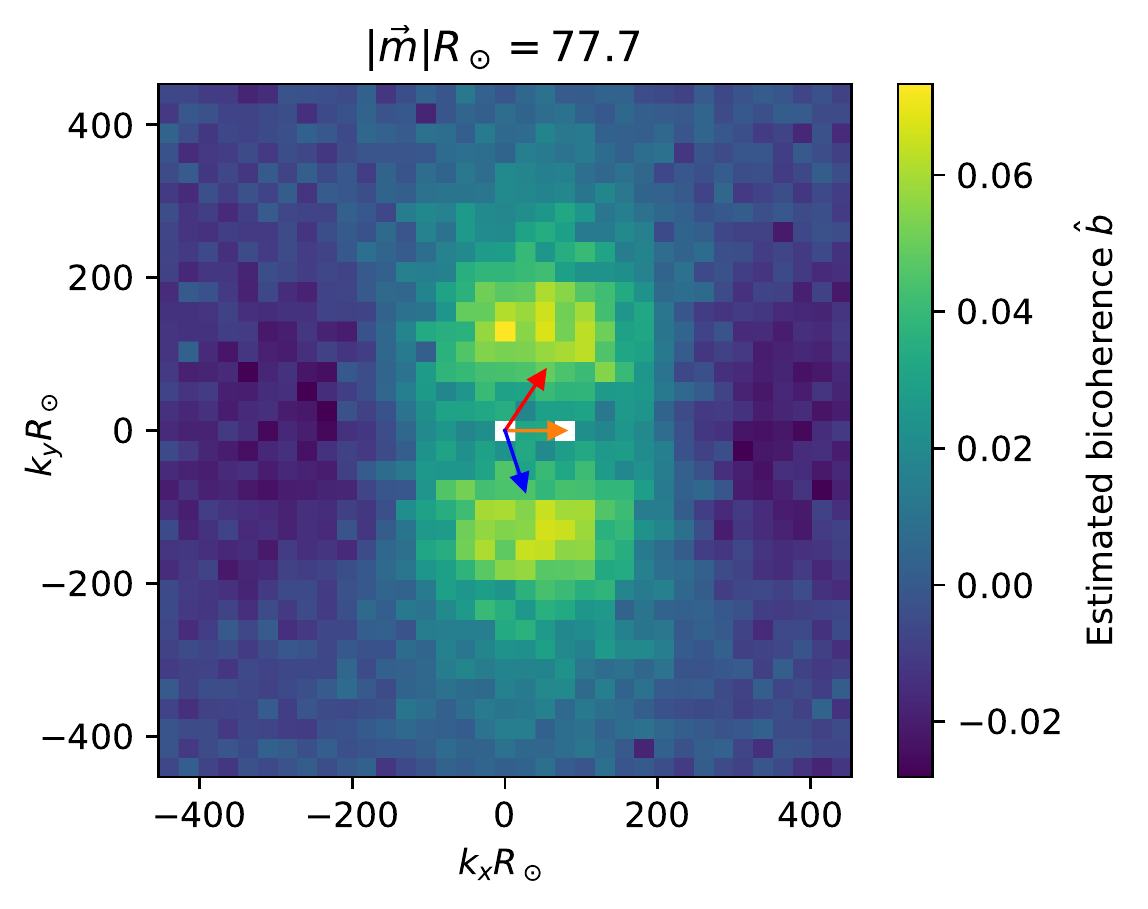}
        \includegraphics[width=0.32\linewidth]{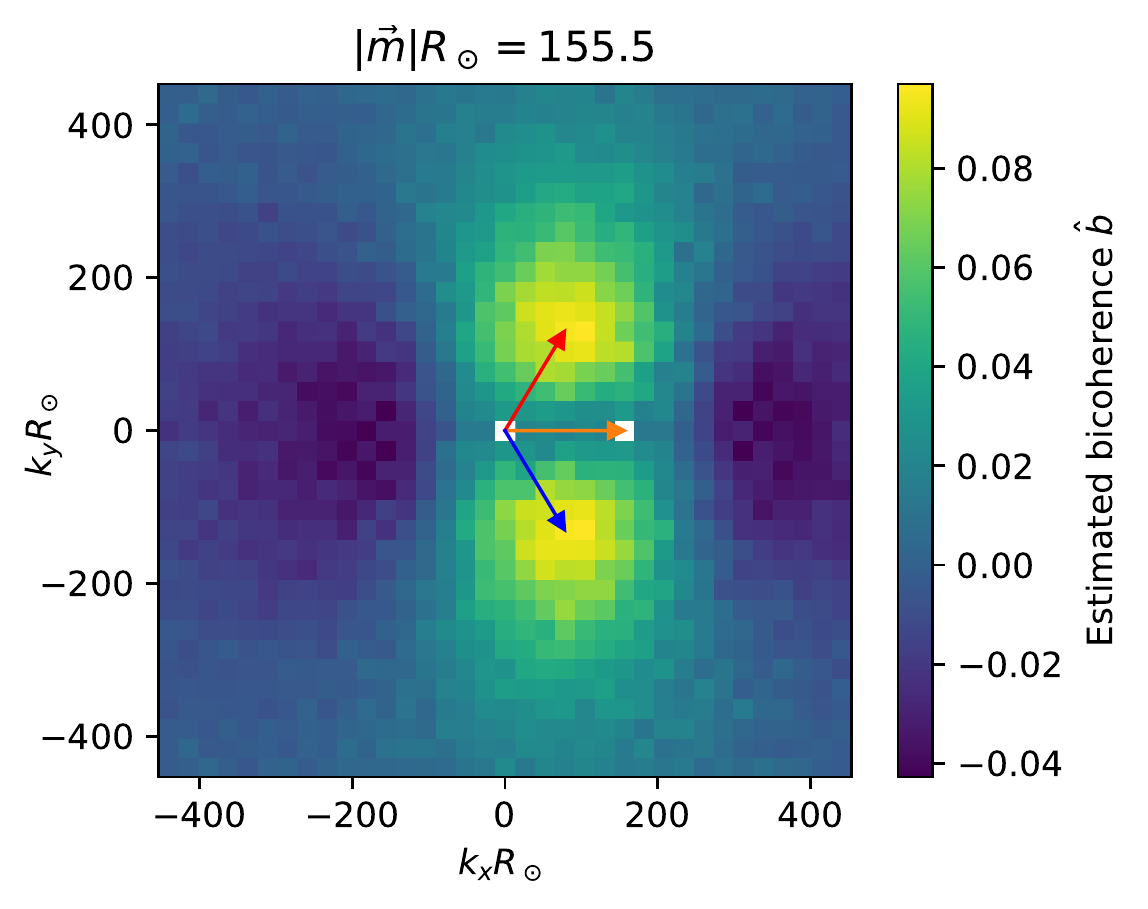}
        \includegraphics[width=0.32\linewidth]{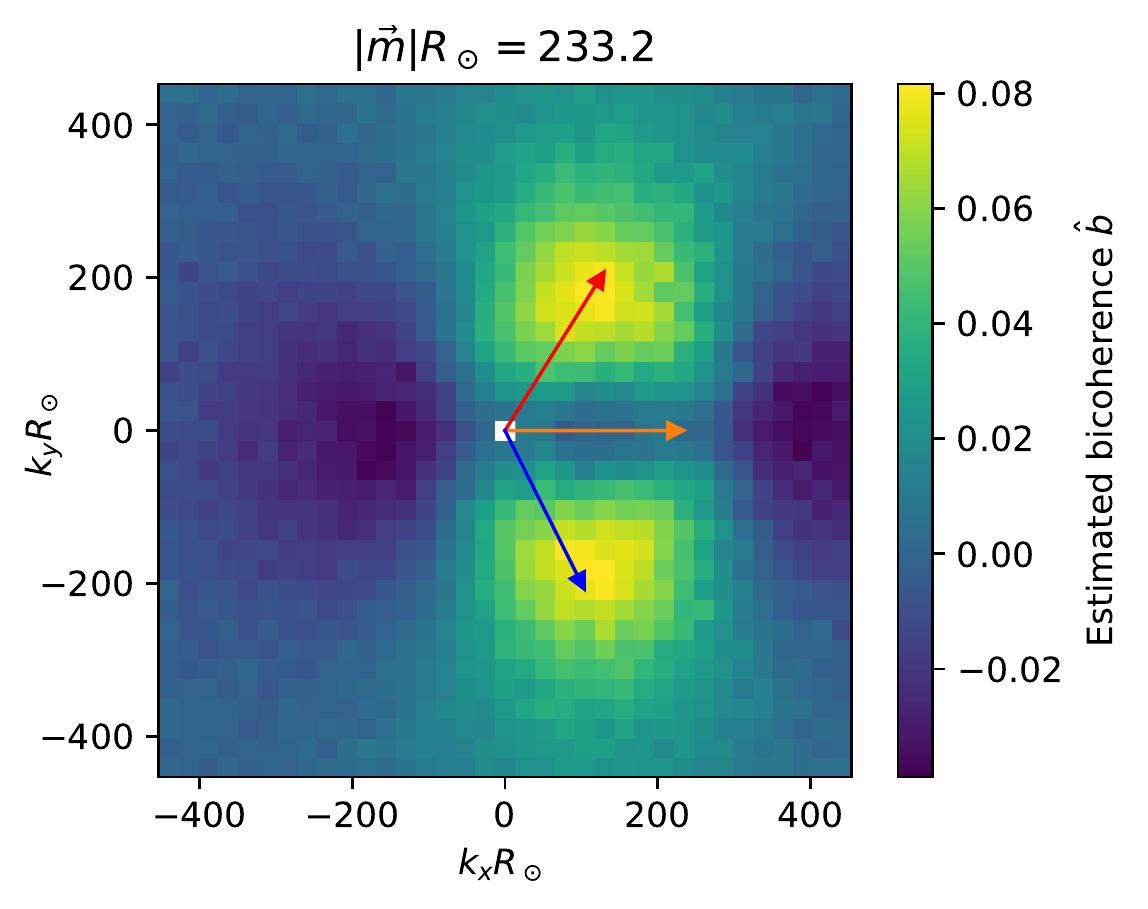}
        \caption{Bispectral coherence $b(\bk,|\bmm|)$ from LCT for selected values of $|\bmm| R_\sun = 77.7, 155.5, 233.2 \pm 12.3$. The vector $\bmm R_\sun$ is fixed for each plot and given by the orange arrow. The bispectral estimates were averaged over the direction of $\bmm$, which is equivalent to assuming that there is no preferred horizontal direction and which results in a real bispectrum. The direction of $\bmm$ in the plot was chosen simply for convenience. The Fourier vectors for a triad $\bk+\bl=\bmm$ that corresponds to hexagons is given by the other two wave vectors (red: $\bk$, {blue}: $\bl$). The angles between the displayed Fourier vectors $\bk$, $\bmm$, and $\bl$ are not exactly 60 degrees because of the discretization. White color denotes locations where no estimate was obtained because one of the three vectors has zero length (as the mean was removed from each map, the bispectral coherence is not defined in this case). See Figure~\ref{figCompBispecBicohDataSynth} for a cut through the middle panel including error bars. See Figures~\ref{figbicohall}-\ref{figbicohKLall} for further illustrations of these results and Figure~\ref{figbicohTD} for the corresponding TD results.}
        \label{figbicoh}
\end{figure*}

Figure~\ref{figbicoh} shows the estimated bicoherence $\hat b$ from LCT. 
The bicoherence is significantly nonzero for a large number of combinations of wave vectors. The bicoherence is largest at wave vectors of similar amplitude, $|\bk|\approx|\bl|$. In general, it is strongest if the coupling vectors are close to the supergranular peak in the power spectrum, in our case at $m R_\sun = 155.5$. At these wave vectors, the amount of power due to a nonlinear component in the data is largest.

Figure~\ref{figbicoh} displays results as a function of $\bk = (k_x,k_y)$ and $|\bmm|$ for selected values of $|\bmm|$. The results are averaged in such a way that $\bmm$ (orange arrow) is rotated to be in the $x$ direction and thereby is fixed; see Appendix~\ref{appBispecEst}. Each data point is then plotted at the location corresponding to $\bk$ relative to the direction of $\bmm$. For each $\bk$, the third coupling Fourier component is then given by $\bl=\bmm-\bk$. The strongest coupling is present at a location where the triad $\bk+\bl = \bmm$ corresponds to a hexagonal pattern at the supergranular scale; see middle panel in Figure~\ref{figbicoh}. Figures~\ref{figbicohall} and~\ref{figbicohKLall} illustrate these results further.

The results for TD are very similar for $|\bmm|R_\sun \lesssim 300$ (Figure~\ref{figbicohTD} and second column in Table~\ref{tabResults}). For $|\bmm|R_\sun > 300$, the signal considerably weakens compared to LCT and is dominated by noise and leakage due to the spatial smoothing. This is the case for both TD datasets.

Estimates for the bispectrum are qualitatively very similar to the bicoherence; see Figure~\ref{figbispec}. As the sign of the bispectrum is predominantly positive, the dominant part of the quadratic interaction results in a positive contribution to the skewness; see Section~\ref{secBispecDef}. This is consistent with the positive skewness of the data; see Figure~\ref{figDataExamples}. The peak in the bispectrum coincides with the peak in the bicoherence. The coupling modes that produce the strongest nonlinearity therefore also result in the largest contribution to the positive skewness in the data.

A positive sign in the bispectral coherence at wave vectors similar to hexagons indicates that the divergence field is preferentially given by outflows surrounded by a network of inflows (see the patterns discussed in Section~\ref{secPatternsIntro}). The outflows are on average slightly stronger while the inflows occupy a larger area. However, in individual realizations of the data, the phase of $X^{(j)}(\bk) X^{(j)}(\bl) X^{(j)*}(\bmm)$ varies substantially. As a consequence, the opposite case of inflows surrounded by outflows is also present in the data, which can be seen in Figure~\ref{figDataExamples}.

In addition, there are regions with negative bispectrum and bicoherence. In those regions, the three vectors $\bk,\bl,\bmm$ are approximatively parallel, with one component being nearly twice as long as the other two. Figure~\ref{figbicohall} shows one example for this in the central feature in the central panel ($|\bmm R_\sun|=310.9$, at $\bk=\bl=\bmm/2$). Due to the symmetry relations $B(\bk,\bl,\bmm)=B(-\bk,\bmm,\bl)^*=B(\bmm,-\bk,\bl)^*$ and as the bispectrum is real in our case, this location has the same bispectral value as a location near the negative peaks in the middle panel in Figure~\ref{figbicoh}. The pattern associated with these negative values and the associated Fourier components is a pattern made of asymmetric rolls, where the regions of outflow are slightly larger than the regions of inflow. This is consistent with the observation that hexagons with a positive bicoherence also have regions of outflow that are larger in diameter than the inflow regions.

\subsection{Generating synthetic data}
\label{secresultsweak}

\begin{figure*}
        \centering

        \includegraphics[width=1\linewidth]{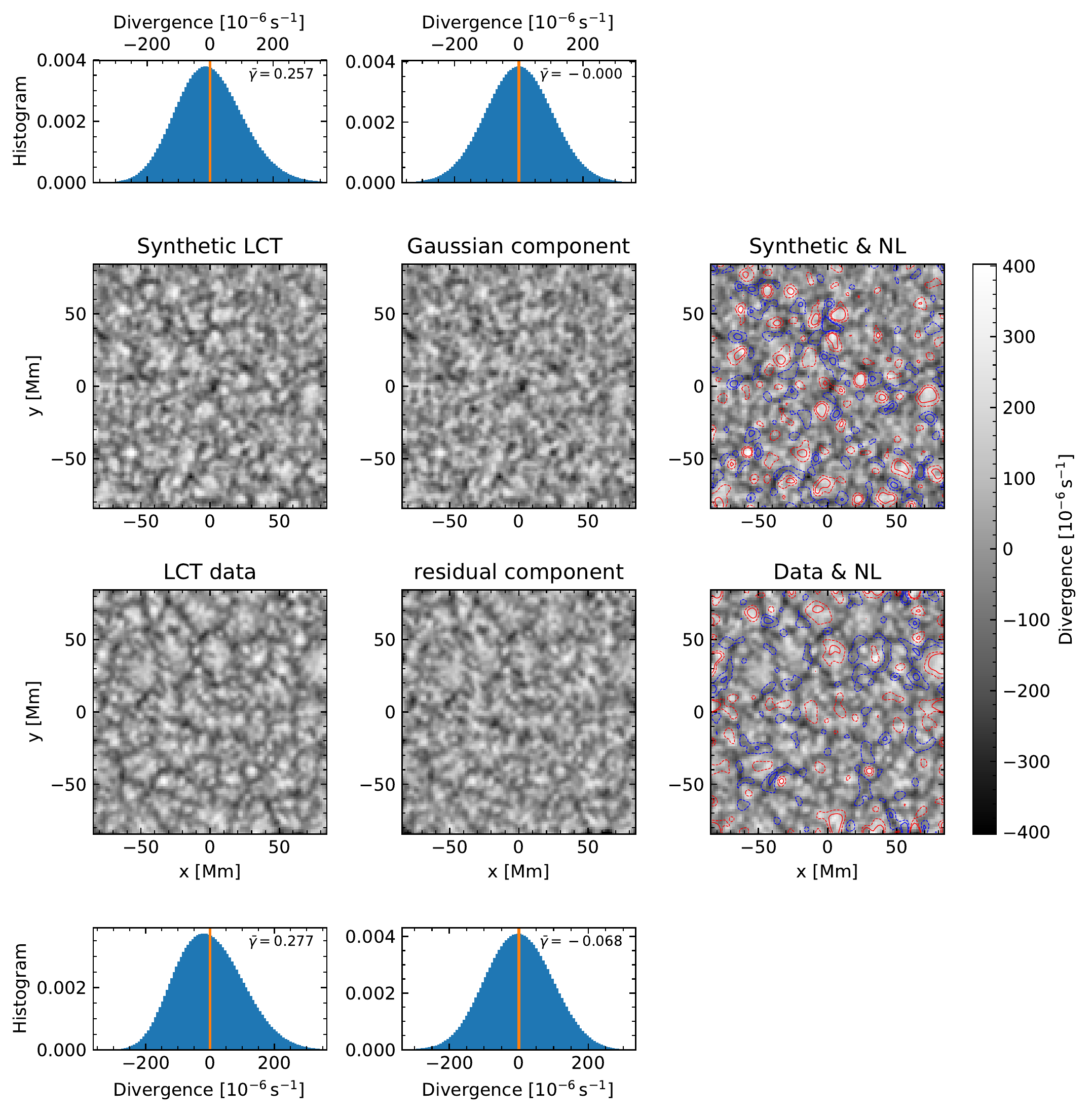}

        \caption{Example synthetic data (top two rows) generated from the measured bispectrum and example LCT data (bottom two rows). The different columns are the (synthetic or real) data (left), {contours of} the {estimated} quadratic nonlinear component overplotted over the data (right), the {residual after subtracting the estimated nonlinear component from the data (middle column, two bottom rows), and the Gaussian field from which the synthetic data were generated (middle column, top two rows)}. Contour lines in the right column indicate levels of $\pm 1\sigma_{\XNL}$ (dashed) and $\pm 2\sigma_{\XNL}$ (solid) of the nonlinear component (positive: red, negative: blue). In addition, the skewness is given in the corresponding histogram; both were computed from the entire dataset. See Figure~\ref{figsynthexampleTD} for the corresponding TD results.}
        \label{figsynthexample}
        \label{figNLest}
\end{figure*}

We generate synthetic data with approximately the given measured power spectrum and bispectrum using the method described in Section~\ref{secgenerate}. As {there is not a unique way to do so, this} is {only} one possible model for the data. Figure~\ref{figsynthexample} shows an example realization of the synthetic LCT data (second row, left column). For comparison, the panel directly below shows an example realization of the LCT data. The synthetic and the real data are very similar and a difference can hardly be seen by eye, apart from the fact that the synthetic data are periodic. Figure~\ref{figsynthexampleTD} shows the corresponding results for TD, which are similar.

\begin{figure*}
        \centering
        \includegraphics[height=0.32\linewidth]{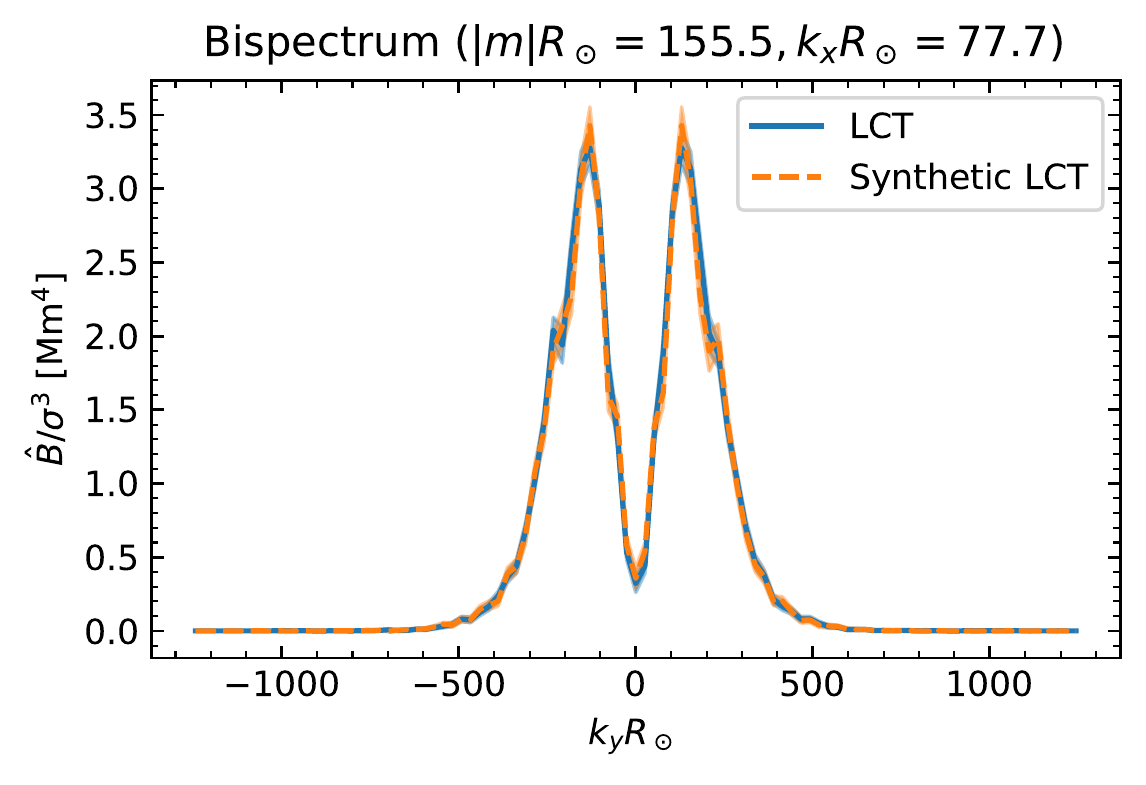}
        \includegraphics[height=0.32\linewidth]{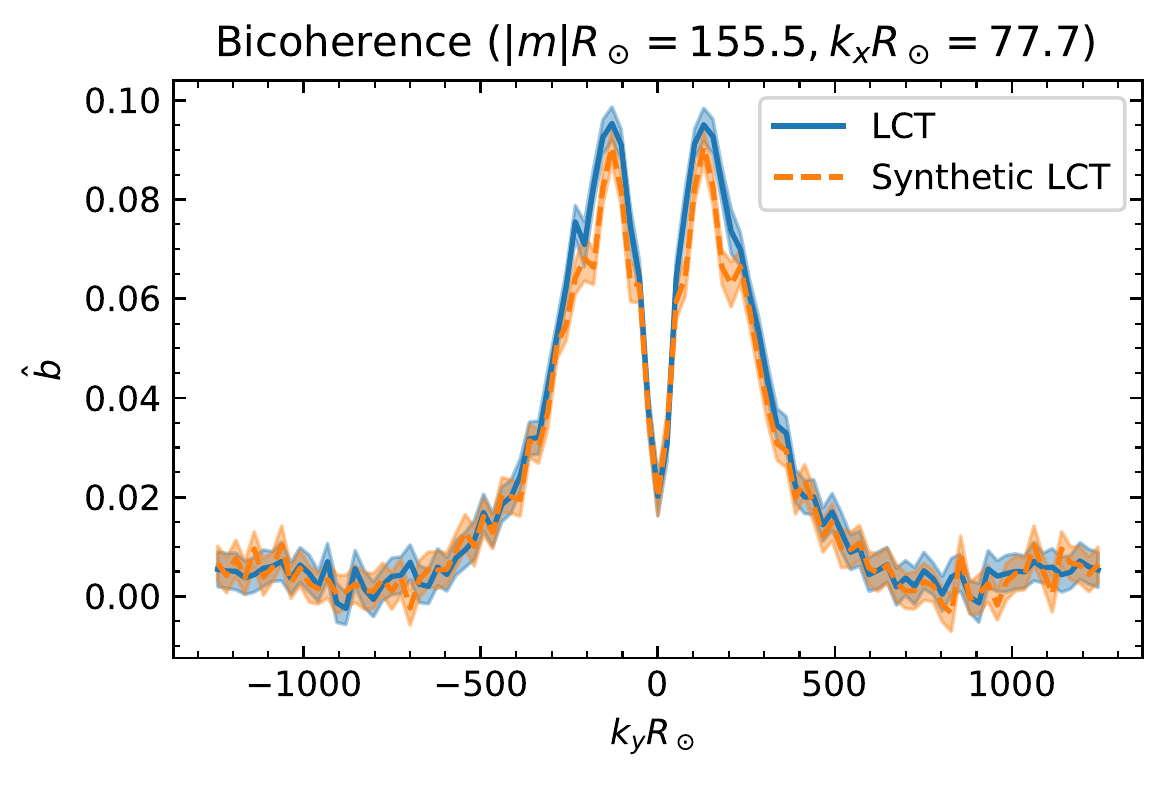}
        \caption{Comparison of estimated bispectrum (left) and bicoherence (right) for real and synthetic LCT data. The shaded regions indicate $1\sigma$ error bars. The LCT data are cuts at $k_x R_\sun = 77.7$ through the middle panels of Figure~\ref{figbispec} (bispectrum) and Figure~\ref{figbicoh} (bicoherence).}
        \label{figCompBispecBicohDataSynth}
\end{figure*}

The bicoherence and bispectrum estimated from the synthetic data agree well with the estimates from the real data (Figure~\ref{figCompBispecBicohDataSynth}) with excellent agreement for the bispectrum. The bicoherence of the synthetic data is slightly lower due to the power in the synthetic data being slightly larger than the real data; see Equation~\eqref{eqvarXgen}, and below for further discussion of this topic.

As we know the nonlinear and Gaussian components of the synthetic data for every realization (e.g., middle and right panels in the top two rows in Figure~\ref{figsynthexample}), we determine the average fraction of the total power (or equivalently of the total variance) stemming from the nonlinear component; see Section~\ref{secNLpower}. We find a value of $\fNL = 4.0\%$, see right{-hand} column of Table~\ref{tabResults}.

\begin{table*}
        \caption{\label{tabResults} Summary of results for the bispectral analysis of supergranular divergence maps from LCT and TD.}
        \centering
        \begin{tabular}{lccccccr}
                \hline\hline
                
                \vspace{0.5mm}

Dataset & $\hatbmax$ &  $\hatmudata$ &  $\hatmub$ &  $\hatmufiltmax$ &  $\hatfNL$ & $\hatfNLmax$  & true $\fNL$ \\
\hline
LCT &  0.097 $\pm$ 0.004 &  0.277 $\pm$ 0.001 &  0.276  &  0.424  &  4.2 \percent &  4.2 \percent &     unknown     \\
Synthetic LCT &  0.091 $\pm$ 0.003 &  0.257 $\pm$ 0.001 &  0.257  &  0.388  &  3.6 \percent &  3.6 \percent &  4.0 \percent \\
Gaussian component (synth. LCT) &  0.028 $\pm$ 0.011 &  -0.000 $\pm$ 0.001 &  -0.000  &  0.003  &  0.0 \percent &  0.0 \percent &  0.0 \percent \\
\hline
TD &  0.099 $\pm$ 0.004 &  0.421 $\pm$ 0.002 &  0.421  &  0.525  &  3.8 \percent &  4.2 \percent &     unknown     \\
Synthetic TD &  0.097 $\pm$ 0.003 &  0.394 $\pm$ 0.002 &  0.393  &  0.500  &  3.3 \percent &  3.8 \percent &  3.6 \percent \\
\hline
Smoothed TD &  0.109 $\pm$ 0.004 &  0.589 $\pm$ 0.003 &  0.589  &  0.595  &  5.6 \percent &  5.6 \percent &     unknown     \\
Synthetic smoothed TD &  0.105 $\pm$ 0.004 &  0.604 $\pm$ 0.004 &  0.603  &  0.608  &  5.4 \percent &  5.4 \percent &  4.9 \percent \\
\hline
\hline
        \end{tabular}
        \tablefoot{For each dataset (left column), the table shows maximal values of the estimated bispectral coherence ($\hatbmax$), estimates of the skewness from the data ($\hatmudata$) and from the bispectrum ($\hatmub$), as well as the maximal value obtainable for low-pass filtered data ($\hatmufiltmax$). By modeling the data with a Gaussian field plus a weak quadratic nonlinear component, we obtain estimates of the fraction of variance due to this nonlinear component ($\hatfNL$), the maximal value obtainable when low-pass filtering the data ($\hatfNLmax$), and the true value of $\fNL$, which is known for the synthetic datasets only. See Figures~\ref{figfnlfilter} and~\ref{figmu3filter} for how $\hatfNL$ and $\hatmub$ depend on low-pass filters applied to the data. The estimate $\hatfNL$ is biased and differs from the true $\fNL$ due to second-order effects, see Eqs.~\eqref{eqfNL} and~\eqref{eqfNLest}. For the smoothed TD data, $\skewData$ was computed here by taking the apodized region in Figure~\ref{figDataExamples} into account to show the consistence with $\hatmub$.}
\end{table*}

In addition, we estimate the nonlinear fraction of power from the bicoherence using Equation~\eqref{eqfNLest}. For the synthetic LCT data,  we estimate a value of $\hatfNL=3.6\%$; see third to last column in Table~\ref{tabResults}. This value is roughly consistent with the true value given the bias introduced by the higher-order term in Equation~\eqref{eqfNL}. This bias is due to the actual power spectrum of the synthetic data being larger than the real data by a factor of $1+\fNL(\bk)$ (see Eq.~\eqref{eqvarXgen}). This reduces the value of the bicoherence by the same amount and results in a small negative bias; see also Figure~\ref{figCompBispecBicohDataSynth}. It is possible to compensate for this effect in future applications by an iterative approach in which the power spectrum of the Gaussian component is successively {adjusted} in amplitude.

\begin{figure}
        \centering
        \includegraphics[width=1\linewidth]{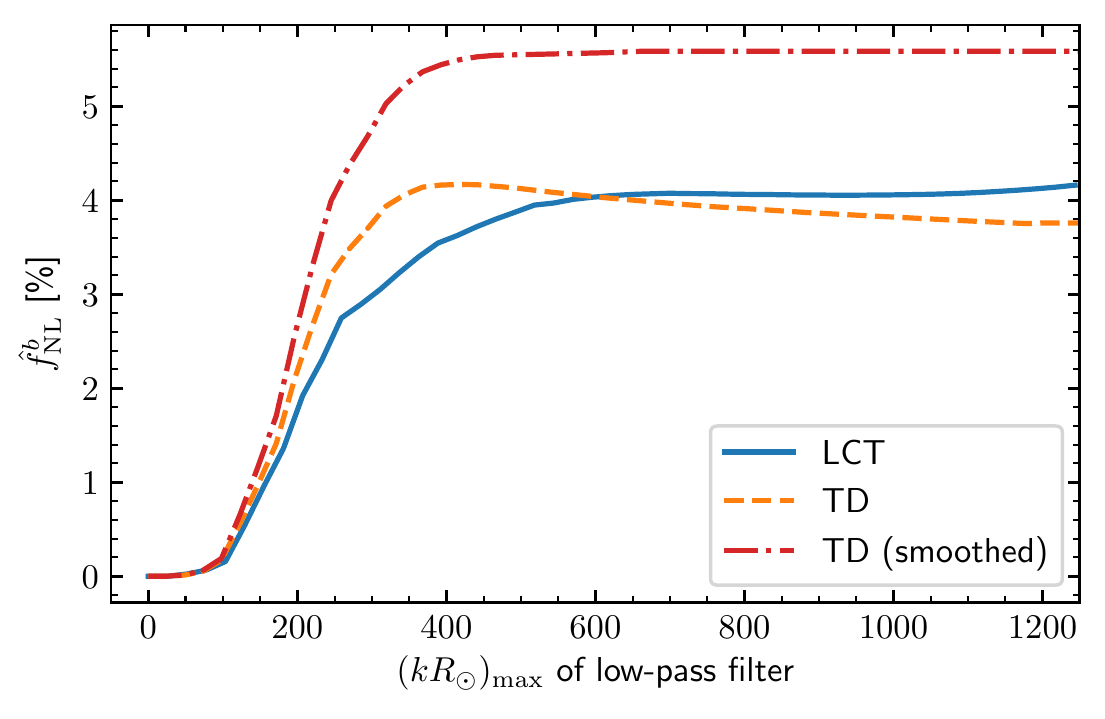}
        \caption{Dependence of $\hatfNL$ on low-pass filter applied to LCT and TD data.}
        \label{figfnlfilter}
\end{figure}

Under the assumption that this model is a good model for our data, we similarly estimate the fraction of quadratic nonlinear power in the data. We obtain values of $4.2\%$ for LCT, $3.8\%$ for TD, and $5.6\%$ for the smoothed TD data (see also Table~\ref{tabResults}). However, this result depends on whether the data are filtered in wave-number space. This is important because both measurements effectively smooth the flow field. To test this, we estimate $\fNL$ for different low-pass filters by limiting the sums over $\bmm$ and $\bk$ in Equation~\eqref{eqfNLest} to $|\bk|,|\bmm|,|\bmm-\bk| \leq k_{\rm{max}}$. Figure~\ref{figfnlfilter} shows the resulting dependence of $\hatfNL$ on $(kR_\sun)_{\rm max}=k_{\rm{max}} R_\sun$. For LCT and the smoothed TD data, the fractional power of the nonlinear component increases until it levels off, while for TD, $\hatfNL$ decreases after attaining its maximum. This is because at large wave numbers, TD is dominated by noise, which was filtered out in the smoothed data. 

\begin{figure}
        \centering
        \includegraphics[width=1\linewidth]{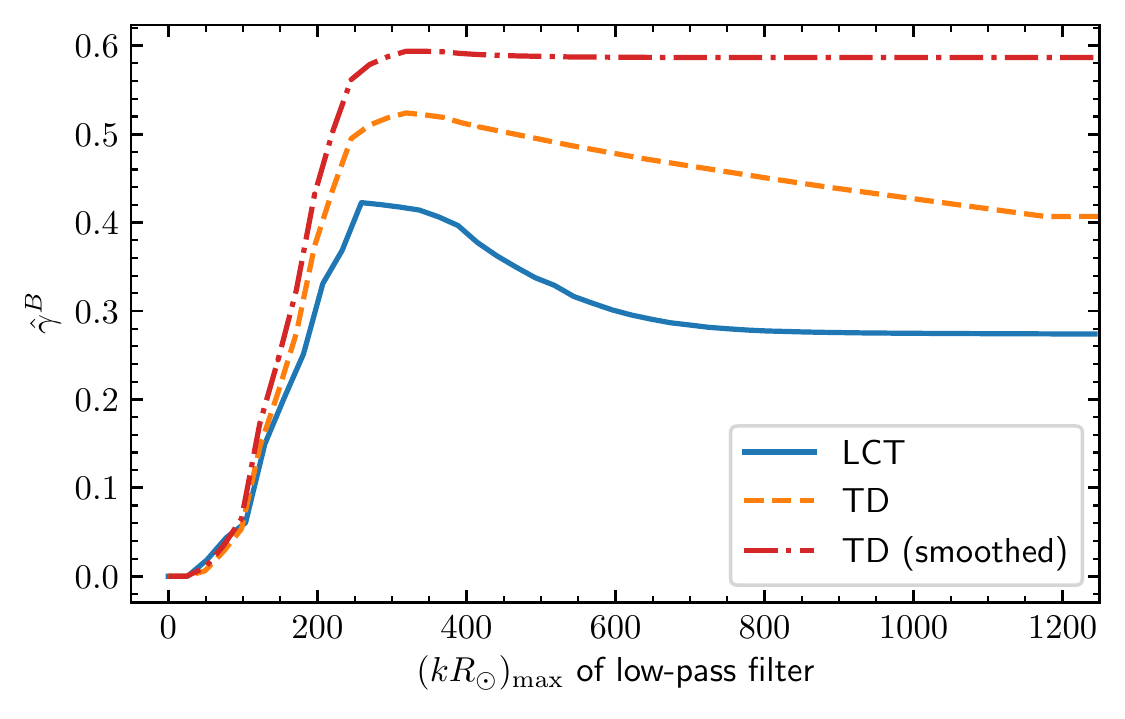}
        \caption{Dependence of  $\hatmub$ on low-pass filter applied to LCT and TD data.}
        \label{figmu3filter}
\end{figure}

In addition to this, we estimate the quadratic nonlinear component directly from the divergence field using Equation~\eqref{eqXNLestimate}. This estimate is biased and may be influenced by higher-order correlations, for example of six Fourier components, which we do not take into account in our study. It therefore serves as an orientation. Figure~\ref{figNLest} (right column) shows how these estimates of the nonlinear component relate to the corresponding realizations of the data. The estimates of the nonlinear component for the real data (right column, bottom panel) and for the synthetic data (right column, top panel) are qualitatively similar. Their relationship to the complete synthetic or real data is similar as well. In both cases, outflows (red contours) are seen in the nonlinear component at locations where an outflow is surrounded by a ring of inflows, or when an inflow is surrounded by a ring of outflows. Inflows (blue contours) are seen mostly in the network between supergranular outflows, either amplifying the network inflow, or suppressing outflows that appear in a network-like location (especially visible in the synthetic case).

Taken together, we conclude that at least 4-6\% of the variance in the data is due to quadratic nonlinear interactions, under the assumption that the data are well explained by a random Gaussian field plus a weak quadratic nonlinear component.

\subsection{Estimating the skewness from the bispectrum}

Finally, we obtain estimates of the skewness from the estimated bispectrum using Equation~\eqref{eqSkewnBispec}; see the fourth column in Table~\ref{tabResults} ($\skewB$) and Appendix~\ref{appBispecEst} for a definition of the estimator. The results are consistent with the estimates from the data, $\skewData$, shown in the third column of Table~\ref{tabResults}. Similar to the fractional power of the nonlinear component, the skewness depends on any low-pass filter applied to the data; see Figure~\ref{figmu3filter}. The estimated skewness of the filtered dataset is obtained using the estimated bispectrum and by limiting the summation in Equation~\eqref{eqSkewnBispec} to $|\bk|,|\bl|,|\bk+\bl| \leq k_{\rm max}  = (kR_\sun)_{\rm max} / R_\sun$ and by estimating $\sigma$ using the similarly filtered power spectrum.

For the synthetic data, the two estimates of the skewness from the data and from the bispectrum agree similarly well; see Table~\ref{tabResults}. However, the skewness in the synthetic data is slightly reduced compared to the real data. This again is due to the total power in the data being larger than the original power by a factor of $1+\fNL(\bk)$ due to second-order effects in the model. This difference could be compensated by an iterative approach in which the power spectrum of the Gaussian component would be adjusted by $1-\fNL(\bk)$. The coupling coefficient $A_{\bk,\bl}$ would have to be adjusted in this case as well to guarantee that the resulting bispectrum is unaltered.

\section{Conclusions}
\label{secConclusion}

We measured the bispectrum and bispectral coherence of the horizontal divergence of near-surface supergranular motions from LCT and TD. The measurements for both datasets agree well and yield significantly nonzero values.

The bispectrum is the next higher-order statistic beyond the power spectrum \citep[e.g.,][Section 11.5.1]{Priestley1981}; it characterizes how the skewness in a dataset is generated by the quadratic interaction of wave vectors that form a triad $\bk+\bl=\bmm$. The bispectral coherence, which is a power-normalized bispectrum, characterizes the strength of a quadratic nonlinear component in the underlying field as a function of the coupling Fourier components, relative to the total power at this Fourier component. 

The bispectral measurements are most significant at supergranular scales, when the three coupling Fourier vectors are of the same magnitude and thus at an angle of about $60\degr$ (middle panel in Figure~\ref{figbicoh}). These Fourier vectors are the same as those of regular hexagons at the supergranular scale. In other words, supergranular scales are the scales where the horizontal divergence pattern looks closest to hexagonal among the scales probed by LCT and TD. At the same Fourier components, the bispectrum is of maximal amplitude and positive, which means that these Fourier components predominantly generate the positive skewness in the dataset (middle panel in Figure~\ref{figbispec}). The positive skewness is due to a preference for broad outflows (positive divergence) surrounded by a weaker network of inflows (negative divergence), which is consistent with the typical appearance of supergranulation (see Figure~\ref{figDataExamples}). The bispectral measurement is consistent with the skewness of the supergranular divergence maps (see Table~\ref{tabResults}).

We use the measured bispectrum to generate synthetic data which look very much like supergranulation (Figures~\ref{figsynthexample} and~\ref{figsynthexampleTD}). The synthetic data have the same bispectrum within error bars and a similar skewness (Figure~\ref{figCompBispecBicohDataSynth} and Table~\ref{tabResults}). To do this, we use a model that consists of a Gaussian field with independent Fourier components plus a weakly nonlinear component, which is generated from the Gaussian one by a quadratic coupling equivalent to the measured bispectrum. This method may be extended in the future to generate synthetic supergranular flow fields. These may be used, for example, for testing helioseismic methods.

Based on this weakly nonlinear model, we find that the nonlinear component is strongest at the supergranular scale. We find that {at least 4-6\%} of the variance is due to this component. This estimate depends on how the small-scale noise is filtered out and it is biased by possibly a few percent (Figure~\ref{figfnlfilter} and Table~\ref{tabResults}). However, for all low-pass filters tested, the estimates are below 6\%.

We find that it is at the supergranular scale that the nonlinear component contributes the largest fraction to the variance. In this sense, the nonlinearities have the strongest effect at supergranular scales. However, the coupling coefficient $A_{\bk,\bl}$ (see Eq.~\eqref{eqAkl}), which is the bispectrum weighted with the inverse of the squared power, increases towards smaller scales.

The bispectral measurements presented in this work may be understood in terms of the example patterns introduced in Section~\ref{secPatternsIntro}. The skewness and the strength of the nonlinear component in the supergranular divergence maps is of similar order to the pattern termed ``small coupling'' (see central panel in the bottom row in Figure~\ref{figPatterns}). This pattern is essentially the same as regular hexagons, but one Fourier component is lowered by a factor of about three.

The relationship between our finding that 4-6\% of the variance is explained by the nonlinear component and the conclusion of \citet{Rincon2017} that supergranular dynamics are strongly nonlinear has yet to be determined. Here we present observations of near-surface flows.
In future work, the depth-dependence of the skewness and the nonlinear component may be studied and the evolution in the time domain may be included in the analysis.

By measuring the bispectrum, we provide a statistical characterization both of the spatial pattern of supergranulation and of the nonlinear interaction in supergranular flows. Together with future studies, this may result in important insight into how supergranulation emerges as a particular length scale on the Sun because it lends itself as a test that simulations and theoretical models of supergranulation 
have to pass. Such a comparison can realistically be made because one needs about 17 days of simulated solar time to measure a bispectral peak as found in this study at a signal-to-noise ratio of about three.

Quadratic nonlinear interactions play an important part in turbulent convective dynamics. This is because the underlying triadic interactions\VB{, which} couple two Fourier modes to a third one\VB{,} are responsible for kinetic energy transport in turbulent flows \citep[e.g.,][]{Alexakis2018}. This may be used in future studies to estimate the direction of the kinetic energy cascade in supergranular flows and may shed light on the still mysterious origin of supergranulation.

\begin{acknowledgements}
We thank Damien Fournier for pointing to a formula that was helpful for the derivations, Friedrich Kupka and Dan Yang for reading and providing feedback on an earlier version of the manuscript, and Michael Wilczek and Friedrich Kupka for helpful discussions. This work was supported in part by the Max Planck Society through a grant on PLATO Science. The computational resources were provided by the German Data Center for SDO through a grant from the German Aerospace Center (DLR). We acknowledge partial support from the European Research Council Synergy Grant WHOLE SUN \#810218.

\end{acknowledgements}



\appendix

\section{Fourier transform convention}
\label{appFourier}

Here we use the same convention for the discrete Fourier transform as \citet{GB2004},
\begin{align}
X(\bk) &= \frac{h_x^2}{(2\pi)^2} \sum_{\bx} X(\bx) \,e^{-i \bk\cdot \bx}, \\
X(\bx) &= h_k^2 \sum_{\bk} X(\bk) \,e^{i \bk\cdot \bx},
\end{align}
where $h_x = L/N_x$ is the sampling interval in one spatial direction, $L$ the spatial extent of the data and $N_x$ the number of samples in one spatial direction. Similarly, $h_k = 2\pi / L$ is the resolution in Fourier space.

\section{Bispectral estimators}
\label{appBispecEst}

We here provide the definitions of the estimators used in this work. Table~\ref{tabFormulas} summarizes the most important quantities and their definitions.

\begin{table*}
        \caption{\label{tabFormulas} Key quantities and definitions used in this paper.}
        \centering
        \begin{tabular}{ccccc}
                \hline\hline
                Symbol & name & definition / formula & see Eq. \# &  major properties\\
                \hline
                {$X(\bx)$} & {random field} & & &  {random variable} \\
                {$X^{(i)}(\bx)$} & {$i$th realization of $X(\bx)$} & & & {single divergence map} \\
        $\sigma^2$ & variance of $X$ & $\sigma^2 = \EE[X(\bx)^2]$ & Eq.~\eqref{eqSigmaDef} & $\EE[X(\bx)]=0$\\
        $\gamma_1$ & skewness of $X$ & $\gamma_1 = \frac{\EE[X(\bx)^3]}{\sigma^{3/2}}$ & Eq.~\eqref{eqSkewnBispec} & \\
        $\skewMap^{(i)}$ & skewness of map $i$& $\skewMap^{(i)} = \frac{\langle X^{(i)}(\bx)^3 \rangle}{\langle X^{(i)}(\bx)^2 \rangle^{3/2}} $ & Eq.~\eqref{eqSigmaDef} & $\langle X^{(i)}(\bx)\rangle = 0 $
        \\
        $\skewData$ & {skewness of dataset} & 
        {$\skewData = \frac{\sum_{i=1}^N \langle X^{(i)}(\bx)^3 \rangle{/N}}{{(}\sum_{i=1}^N \langle X^{(i)}(\bx)^2 \rangle{/N}{)}^{3/2}}$} &  & \\
        $B$ & bispectrum\tablefootmark{a} & $B(\bk,\bl) = h_k^2 \EE[X(\bk)X(\bl)X^*(\bk+\bl)]$ & Eq.~\eqref{eqBispecDiscrete} & $\gamma_1 = \sum_{\bk,\bl} B(\bk,\bl) \, h_k^4\, / \,\sigma^{3}$ \\
        $P$ & power spectrum\tablefootmark{a} & $P(\bk)=h_k^2 \EE[|X(\bk)|^2]$ & Eq.~\eqref{eqPowerDef} & $\sigma^2 = \sum_\bk P(\bk) \, h_k^2$ \\
        $b$ & bicoherence\tablefootmark{b} & $b(\bk,\bl) = \frac{\EE[X(\bk)X(\bl)X^*(\bk+\bl)]}{ \sqrt{\EE[|X(\bk)X(\bl)|^2]\EE[|X(\bk+\bl)|^2]}}$ & Eq.~\eqref{eqbDef} & $\hat B$ significant if $|\hat b| > \sigma_b = N_B^{-0.5}$ \\
        $N$ & \# of realizations of $X$ & & & $N=365\times11=4015$ \\
        $N_B$ & \# of realiz. used in $\hat B$ & $N_B(\bk,\bl)\approx \pi \frac{|\bk+\bl|}{h_k} N$ & & $N_B > N$ due to azim. average\\
        $A_{\bk,\bl}$ & coupling coefficient & $A_{\bk,\bl} = \frac{1}{6} \frac{h_k^2\, B(\bk,\bl) }{P(\bk)P(\bl)}$ & Eq.~\eqref{eqAkl} & \\
        $X_{\rm NL}$ & NL\tablefootmark{c} model component & $X_{\rm NL}(\bmm) = \sum_{\bk} A_{\bk, \bmm -\bk} X_{\rm G}(\bk) X_{\rm G}(\bmm -\bk)$ & Eqs.~\eqref{eqXgenerate}-\eqref{eqXNLgenerate} & $X(\bmm) = X_{\rm G}(\bmm) + X_{\rm NL}(\bmm)$ \\
        $\fNL(\bk)$ & NL\tablefootmark{c} fraction of $P(\bk)$ & $\fNL(\bk) = \frac{\EE[|X_{\rm NL}(\bk)|^2]}{\EE[|X_{\rm G}(\bk)|^2]}$ & Eqs.~\eqref{eqEXNLm2},\eqref{eqfNLm} & \\
        $b_P$ & bicoherence ($P$-norm.) & $b_P(\bk,\bl) = \frac{h_k B(\bk,\bl)}{\sqrt{P(\bk) P(\bl) P(\bmm)}}$ & Eqs.~\eqref{eqbPDef},\eqref{eqfNLm} & $\fNL(\bmm) = \frac{1}{18} \sum_{\bk} b_P(\bk,\bmm-\bk)^2 $ \\
        $\fNL$ & NL\tablefootmark{c} fraction of {$\sigma^2$} & $\fNL = \frac{\EE[|X_{\rm NL}(\bx)|^2]}{\EE[|X(\bx)|^2]}$ & Eq.~\eqref{eqfNL} & \\
        \hline
        \end{tabular}
        \tablefoot{Estimators for these quantities are given by averaging realizations for estimating the expectation values in the above formulas; see Appendix~\ref{appBispecEst}. $\langle f(\bx) \rangle$ denotes a spatial average of $f$. The data have zero spatial mean by construction. \tablefoottext{a}{The bispectrum and power spectrum as defined here are actually bispectral and power spectral densities.}
                \tablefoottext{b}{Also known as bispectral coherence.}
                \tablefoottext{c}{NL is shorthand for nonlinear.}
        }
\end{table*}

\subsection{Azimuthal averaging}
\label{appAzimAvg}

Assuming that supergranulation at the equator does not have a preferred horizontal direction, it is possible to average the bispectral estimates over circles with the same (or similar) values of $|\bmm|$. This reduces the noise in the measurement and can be done as follows. For a given triad  $\bmm=\bk+\bl$, all three vectors are rotated clockwise by the angle between $\bmm$ and $\tilde \bmm = (|\bmm|,0)$ to obtain rotated vectors $\tilde \bmm = \tilde \bk + \tilde \bl$. The resulting vector $\tilde \bmm$ then points in the $x$-direction and the realization $X^{(j)}(\bk) X^{(j)}(\bl) X^{(j)}(\bmm)^*$ is taken as a realization of $X^{(j)}(\tilde \bk) X^{(j)}(\tilde \bl) X^{(j)}(\tilde \bmm)^*$ in the averaging procedure.

Equivalently, it is possible to simply average the estimates for the bispectrum in the same way. This reduces the bispectral estimates from four-dimensional to three-dimensional variables, which can be expressed either as a function of $(\bk,|\bmm|)$ or $(|\bk|,|\bl|,|\bmm|)$.

We have to take into account that Fourier components of opposite sign have identical data up to complex conjugation and therefore do not add any new information. The number of relizations of $X^{(j)}$ is $N=4015$ and the number of realizations used at $\bmm$ is then $N_B=N_{|\bmm|} N$, where $N_{|\bmm|}\approx \pi \frac{|\bk+\bl|}{h_k}$ is half of the number of vectors $\tilde \bmm$ with the same length as $\bmm$, that is the number of elements of the set
\begin{align}
M_{|\bmm|} = \{\tilde \bmm \,\textrm{such that:}\, |\bmm|-\frac{h_k}{2} \leq |\tilde \bmm| <  |\bmm|+\frac{h_k}{2} \}.
\end{align}

The azimuthal averaging leads to a different number of realizations being used for different values of $|\bmm|$ so that the errors in the estimates of the bicoherence and bispectrum depend on $|\bmm|$. In addition, this number varies slightly with the value of $\bk$ due to the discretization and due to rounding. For example, at the supergranular peak in the power spectrum, which in our case is at $|\bmm| R_\sun=155.5$, we averaged over a typical number of $N=80300$ realizations. This leads to a statistical error for the bispectral coherence of about $\sigma_b=0.0035$ at $|\bmm| R_\sun=155.5$.

\subsection{Definition of bispectral estimators}

To shorten notation, we write
\begin{align}
    \sum = \sum_{\tilde \bmm\in M_{|\bmm|}}  \sum_{j=1}^N.
\end{align}
{Following \cite{Kim1979}}, our estimate of the bispectrum, $\hat B$, can then be written
\begin{align}
\hat B(\bk,|\bmm|) &= h_k^2 \,\frac{1}{N_B} \sum X^{(j)}(\tilde \bk) X^{(j)}(\tilde \bl) X^{(j)}(\tilde \bmm)^* , \label{eqBispecEstimate}
\end{align}
where the vectors $\tilde \bk,\tilde \bl,\tilde \bmm$ are obtained from $\bk,\bl=\bk-\bmm,\bm$ by a rotation of $-\arcsin{\frac{m_x}{|\bmm|}}$, see Appendix~\ref{appAzimAvg}.

Our estimate for the bicoherence then becomes
\begin{align}
\hat b(\bk,|\bmm|) &= \frac{\frac{1}{N_B} \sum X^{(j)}(\tilde \bk) X^{(j)}(\tilde \bl) X^{(j)}(\tilde \bmm)^*}{\sqrt{\frac{1}{N_B}\sum |X^{(j)}(\tilde\bk) X^{(j)}(\tilde \bl)|^2 \frac{1}{N_B}\sum |X^{(j)}(\tilde\bmm)|^2}} , \label{eqBicohEstimate}
\end{align}
and similarly for the power-normalized bicoherence,
\begin{align}
\hat b_P(\bk,|\bmm|) &= \frac{ \frac{1}{N_B} \sum X^{(j)}(\tilde \bk) X^{(j)}(\tilde \bl) X^{(j)}(\tilde \bmm)^*}{\sqrt{\frac{1}{N_B}\sum |X^{(j)}(\tilde\bk)|^2 \frac{1}{N_B}\sum |X^{(j)}(\tilde \bl)|^2 \frac{1}{N_B}\sum |X^{(j)}(\tilde\bmm)|^2 }}. \label{eqBicohPEstimate}
\end{align}
As a consequence, one has
\begin{align}
    \hat B(\bk,\bl) &= \hat B(\tilde \bk,|\tilde \bmm|) \\
    \hat b(\bk,\bl) &= \hat b(\tilde \bk,|\tilde \bmm|) \\
    \hat b_P(\bk,\bl) &= \hat b_P(\tilde \bk,|\tilde \bmm|).
\end{align}

\section{Details on the weakly nonlinear model}
\label{appModel}

We here show the basic steps for demonstrating that the bispectrum of the synthetic data generated with the weakly nonlinear model of Section~\ref{secgenerate} is the same as the input bispectrum to leading order.

To do that, one first observes that different Fourier components of the nonlinear {component} are uncorrelated,
\begin{align}
\EE[X_{\rm NL}^*(\bmm) X_{\rm NL}(\bmm')] &=\delta_{\bmm,\bmm'} \EE[|X_{\rm NL}(\bmm)|^2],
\end{align}
{where $\delta_{\bmm,\bmm'}$ is the Kronecker symbol and the nonlinear component is uncorrelated with the Gaussian component,
\begin{align}
\EE[X_{\rm NL}^*(\bmm) X_{\rm G}(\bmm')] &= 0.
\end{align}}

Consequently, the total power is proportional to
\begin{align}
\EE[X^*(\bk) X(\bl)] &= \delta_{\bk,\bl} \, \left( \EE[|X_{\rm G}(\bk)|^2] +  \EE[|X_{\rm NL}(\bk)|^2] \right) \\
&= \frac{1}{h_k^2} \delta_{\bk,\bl} P(\bk) \Big( 1 + \fNL(\bk)  \Big),\label{eqvarXgen}
\end{align}
where $\fNL(\bk)$ is the ratio of then nonlinear power with regard to the background power of the Gaussian component at $\bk$,
\begin{align}
\fNL(\bk)  &= \frac{\EE[|X_{\rm NL}(\bk)|^2]}{\EE[|X_{\rm G}(\bk)|^2]}  =  O(|X_{\rm NL}|^2), \label{eqfNLk}
\end{align}
and where $O(|X_{\rm NL}|^n)$ denotes terms that are of $n$th order in the nonlinear component.

The bispectrum of $X$ is then \citep[see also][]{Wagner2010}
\begin{align}
B(\bk,\bl) &= \frac{2\delta_{\bk+\bl,\bmm}}{h_k^2} \Big[ P(\bk)P(\bl) A_{\bk,\bl} +  P(\bk)P(\bmm) A_{\bk,-\bmm} \nonumber\\
&\quad + P(\bl)P(\bmm)  A_{\bl,-\bmm}\Big]  +  O\Big(|X_{\rm NL}|^3\Big). \label{eqBispecAkl} 
\end{align}

Using Eq.~\eqref{eqAkl} and using the symmetry relation $B_0(\bk,\bl,{\bmm})=B_0(\bk,-\bmm,-\bl)=B_0(\bl,-\bmm,-\bk)$, which is fulfilled by every bispectrum, we then obtain
\begin{align}
B(\bk,\bl) &= B_0(\bk,\bl)  +  O\Big(|X_{\rm NL}|^3\Big), \label{eqBispecB0}
\end{align}
where the correction term is given by
\begin{align}
O\Big(|X_{\rm NL}|^3\Big) = h_k^2 \, \EE[X_{\rm NL}(\bk) X_{\rm NL}(\bl) X_{\rm NL}^*(\bmm)].
\end{align}
If the nonlinearity is weak, the term $O(|X_{\rm NL}|^3)$ can be neglected and this model is suited to generate data with a given bispectrum.

If we generate synthetic data for a bispectrum $B_0=\hat B$, which was estimated from the data, we set $A_{\bk,\bl}$ to zero if the estimated bicoherence $\hat b$ is not significantly different from zero. Using too many nonsignificant coupling terms results in additional random nonlinear terms in the data. In this work, we use a threshold for $|\hat b|$ of about $2\sigma_b$ ($1.7\sigma_b$ for LCT and $2\sigma_b$ for TD). This threshold was chosen to balance the removal of noise with keeping the signal. This procedure is in general agreement with the discussion in \cite{Kim1979}.

\section{Estimating {the power of} the nonlinear {component}}
\label{secNLpower}

Given the above model, one can estimate the fraction of power due to the nonlinear interaction based on the bispectral estimates. To do that, one considers the power at a Fourier component $\bmm$,
\begin{align}
\frac{\EE[|X_{\rm NL}(\bmm)|^2]}{\EE[|X(\bmm)|^2]} = \frac{\fNL(\bmm)}{1+\fNL(\bmm)} = \fNL(\bmm) + O \left(\fNL(\bmm)^2 \right), 
\label{eqEXNLm2}
\end{align}
and finds
\begin{align}
\fNL(\bmm)  &= \frac{1}{18} \sum_{\bk} b_P(\bk,\bmm-\bk)^2 \label{eqfNLm},
\end{align}
where $b_P$ is similar to the bicoherence, but normalized by the power spectra (compare to Eq.~\eqref{eqbDef}),
\begin{align}
b_P(\bk,\bl) &= \frac{h_k B_0(\bk,\bl)}{\sqrt{P(\bk) P(\bl) P(\bmm)}} \label{eqbPDef} \\
&= \sqrt{1+\delta_{\bk,\bl}} \, b(\bk,\bl) + \textrm{higher order terms}.
\end{align}

To first order, the fraction of power of the nonlinear component at $\bmm$ is thus proportional to the squared bicoherence, $b^2$, summed over all triads that add up to $\bmm$ (except at $\bk=\bl$). The factor of $1/18$ stems from the multiple permutations that are present when computing expectation values of four multiplied Gaussians in Equations \eqref{eqBispecAkl} and \eqref{eqfNLm}.

The fractional variance from the nonlinear component among the total field can similarly be computed,
\begin{align}
\fNL&= \frac{\EE[|X_{\rm NL}|^2]}{\EE[|X|^2]} 
&= \sum_\bmm \fNL(\bmm) \frac{ P(\bmm)}{\sum_{\bmm'} P(\bmm')} + O \left( |X_{\rm NL}|^2 \right)
.  \label{eqfNL}
\end{align}
The fraction of the total variance that is due to the nonlinear component, $\fNL$, is thus a power-weighted average of the fractional nonlinear power at $\bmm$, plus a small higher-order correction. We thus estimate the fraction of power due to the nonlinear {component} by 
\begin{align}
 \hatfNL &= \frac{1}{18} \sum_{\bmm,\bk} |\hat b_P(\bk,\bmm-\bk)|^2 \frac{ P(\bmm)}{\sum_{\bmm'} P(\bmm')}, \label{eqfNLest}
\end{align}
where $\hat b_P$ is given in Appendix~\ref{appBispecEst}. From Equations~\eqref{eqfNL} and~\eqref{eqfNLest}, it is clear that $\hatfNL$ is biased due to the higher-order correction in Equation~\eqref{eqfNL}. When computing the sum in Equation~\eqref{eqfNLest}, we again only keep terms where the bicoherence $\hat b$ is significantly nonzero by at least $2\sigma_b$. This is consistent with using the same threshold for generating synthetic data as discussed at the end of Appendix~\ref{appModel}.

{The bispectrum $B$ and its variance-normalized version $B/\sigma^3$ are effectively bispectral densities (see Eq.~\eqref{eqSkewnBispec}, which includes an $h_k^4$ term). As such, they are in principle invariant under changes in the box size, once the resolution of the Fourier grid is much smaller than the scale on which $B$ varies. The same is not true for the bicoherence (Eqs.~\eqref{eqfNLm} and~\eqref{eqfNL} do not include an $h_k^4$ term).}

\section{Estimating the nonlinear component}

\label{secNL}

In addition, it is possible to estimate the nonlinear component of the {divergence} field at each realization. One starts from Equation~\eqref{eqXNLgenerate} and simply inserts the data $X$ instead of the Gaussian component,
\begin{align}
\hat X_{\rm NL}^{(j)}(\bmm) &= \sum_{\bk} A_{\bk, \bmm -\bk} X^{(j)}(\bk) X^{(j)}(\bmm -\bk) \label{eqXNLestimate},
\end{align}
where $A_{\bk, \bmm -\bk}$ is obtained from the estimated bispectrum $\hat B$ and Equation~\eqref{eqAkl} with $B_0 = \hat B$. Again, we keep only terms where the estimated bicoherence $\hat b$ is significantly nonzero. The estimator $\hat X_{\rm NL}$ fulfills
\begin{align}
\hat X_{\rm NL}^{(j)}(\bmm) &= X_{\rm NL}^{(j)}(\bmm) + O(|X_{\rm NL}^{(j)}|^2).
\end{align}
We note that $\hat X_{\rm NL}$ is a biased estimator.

\section{Additional figures}
\label{appAddFigs}

To further illustrate our results from Section~\ref{secresults}, this appendix includes a few additional figures.

\subsection{Results for TD}

\begin{figure*}
        \centering
        \includegraphics[width=0.32\linewidth]{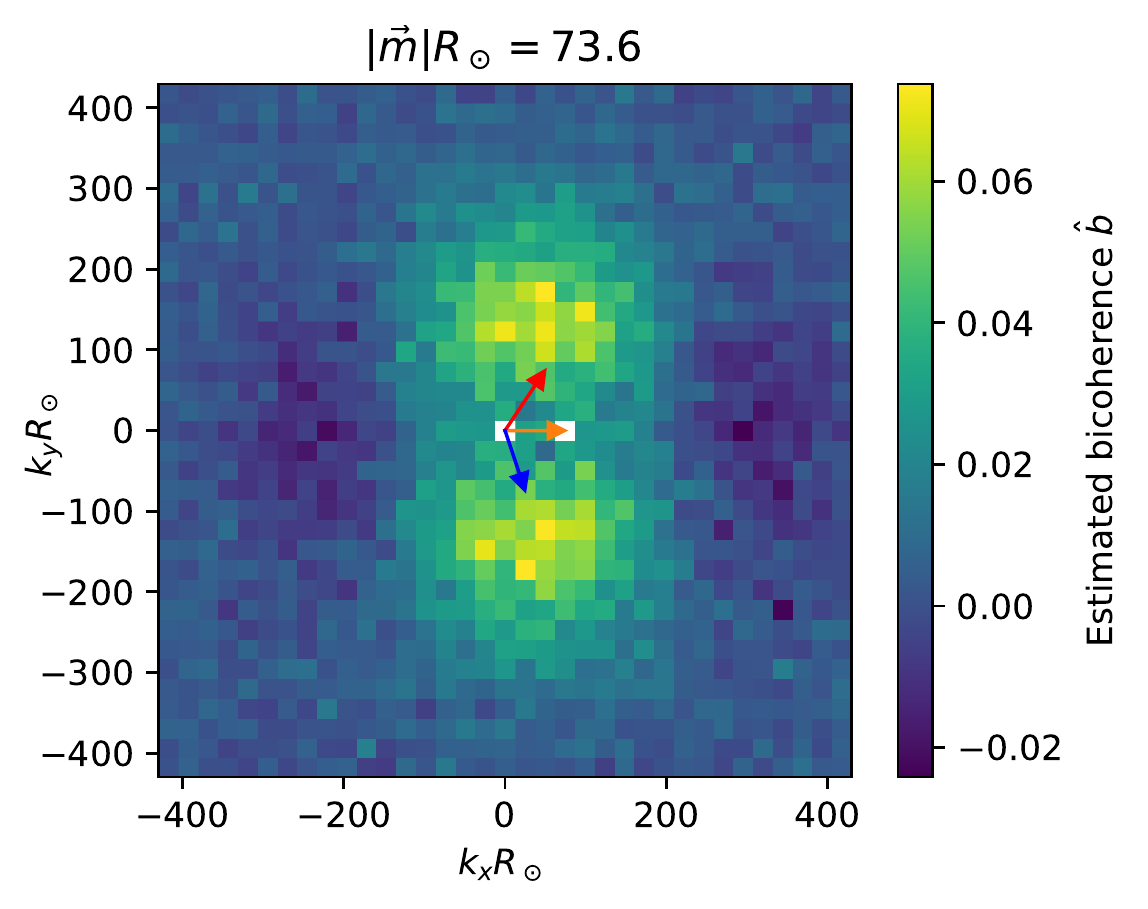}
        \includegraphics[width=0.32\linewidth]{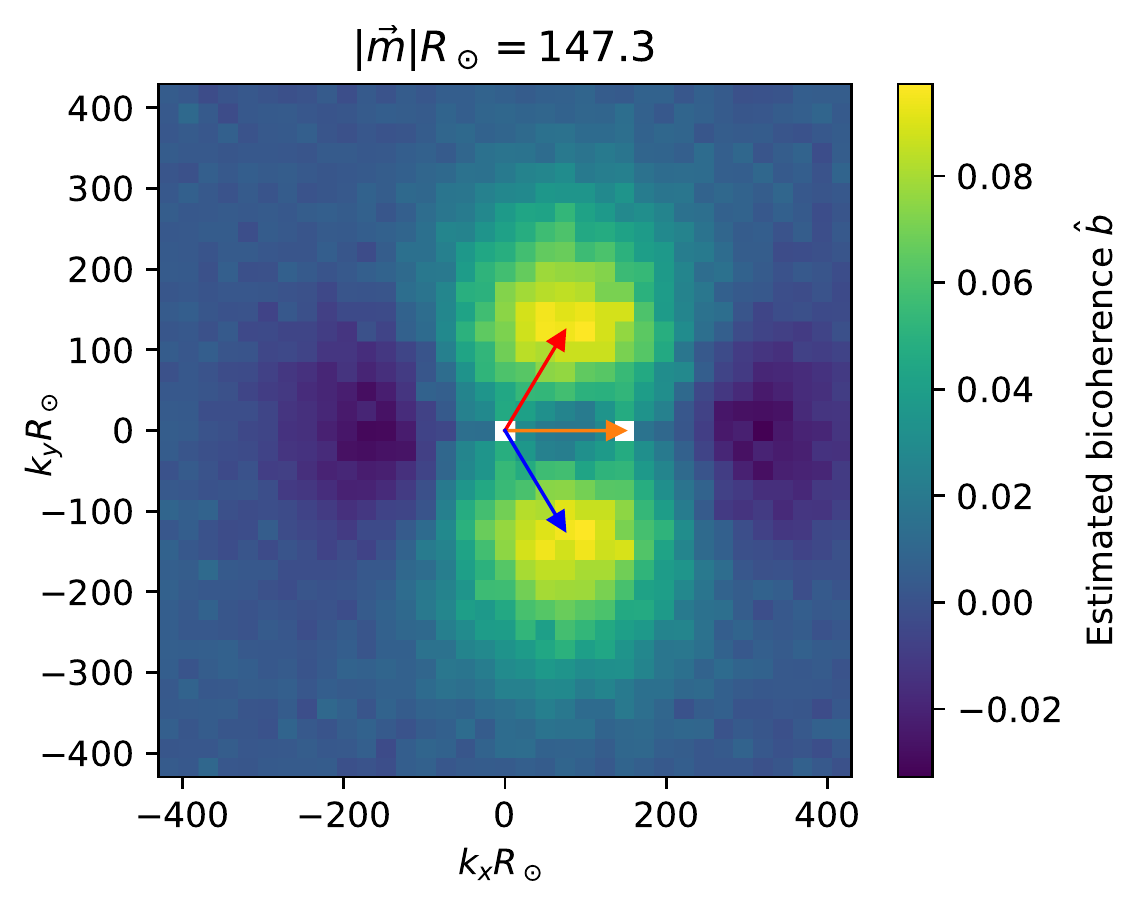}
        \includegraphics[width=0.32\linewidth]{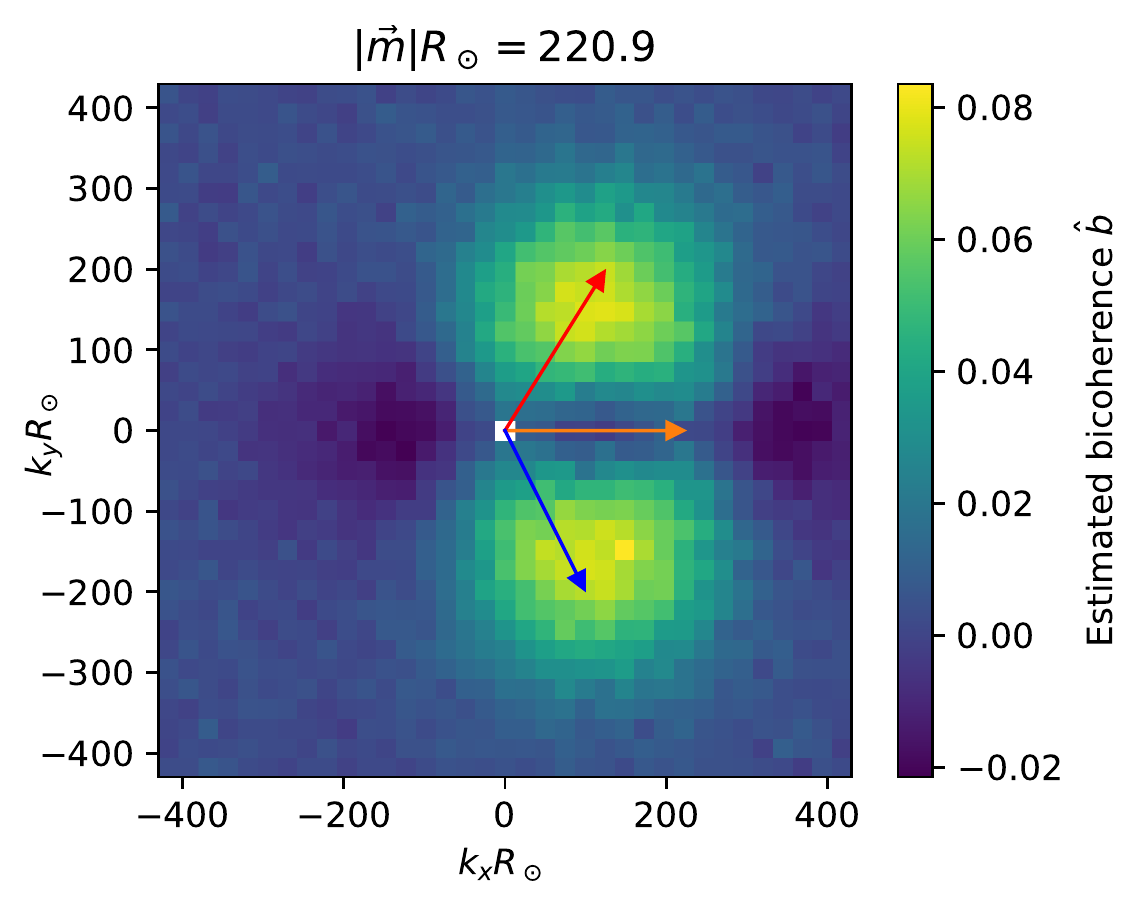}
        \caption{Same as Figure~\ref{figbicoh}, but for TD and $|\bmm| R_\sun = 73.6 , 147.3 , 220.9 \pm 13.0$.}
        \label{figbicohTD}
\end{figure*}

Figure~\ref{figbicohTD} is essentially the same as Figure~\ref{figbicoh} and shows results for the bicoherence $b$ for the (unsmoothed) TD data for selected values of $|\bmm| R_\sun$ that are closest to those in Figure~\ref{figbicoh}. Results for the smoothed TD dataset are not shown but very similar with slightly less noise and slightly larger values, which is due to the apodization. Both TD results compare well with LCT.
Figure~\ref{figsynthexampleTD} shows the same comparison of synthetic and real data for the unsmoothed TD dataset that Figure~\ref{figNLest} shows for LCT.

\subsection{Further details on LCT results}

\begin{figure*}
        \centering
        \includegraphics[width=0.32\linewidth]{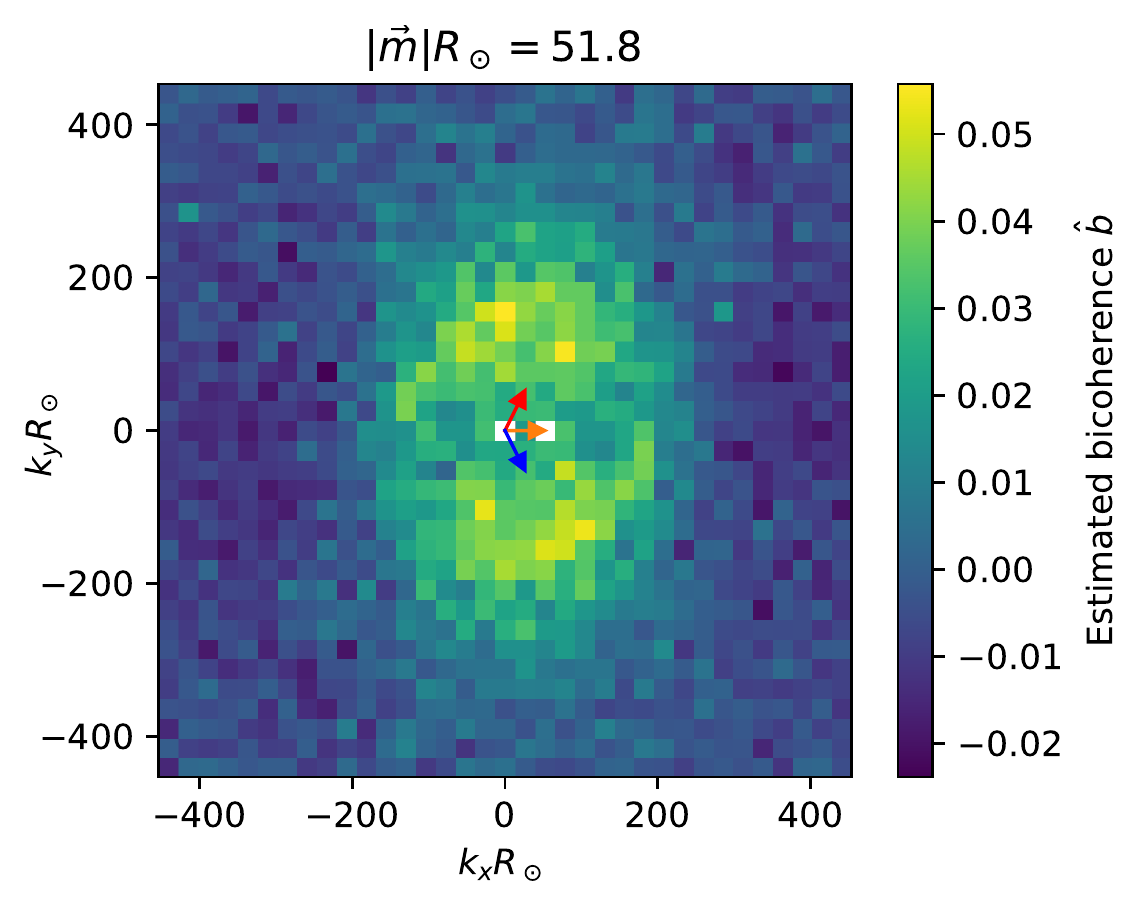}
        \includegraphics[width=0.32\linewidth]{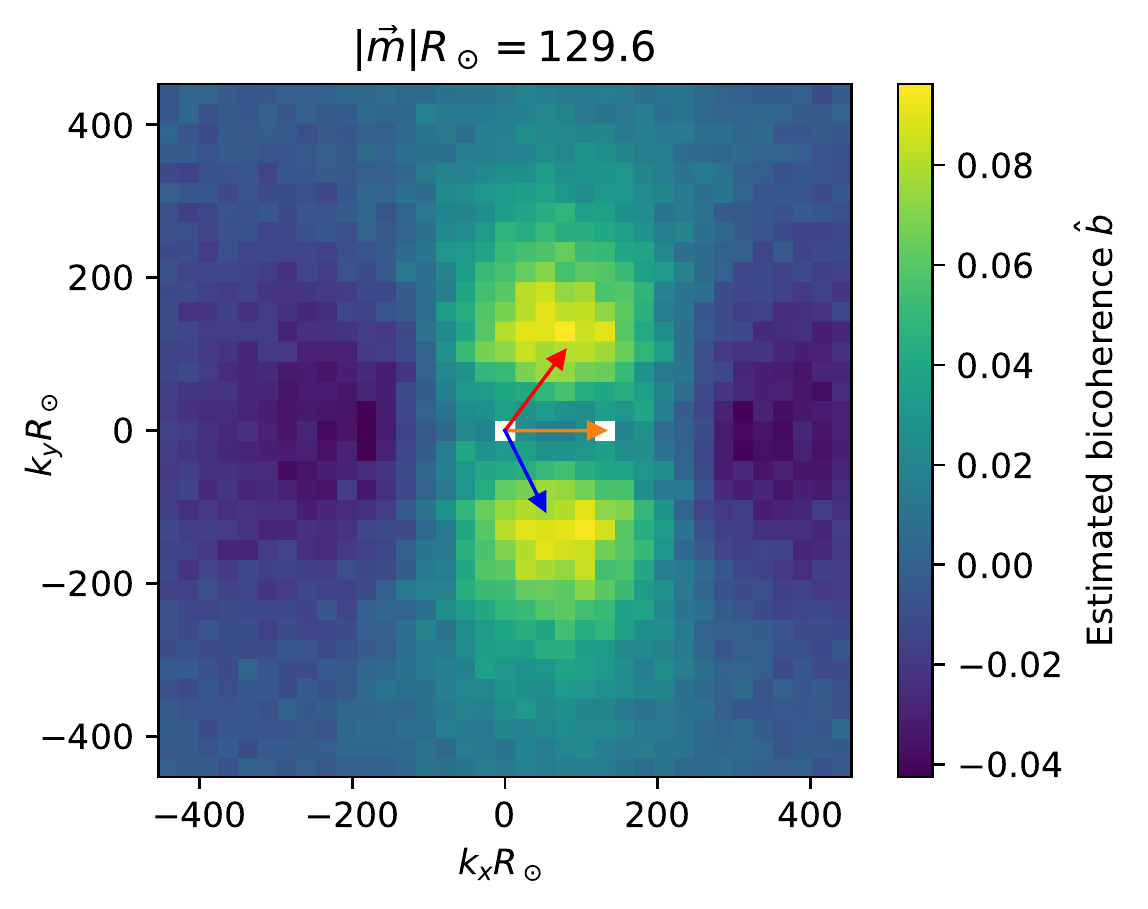}
        \includegraphics[width=0.32\linewidth]{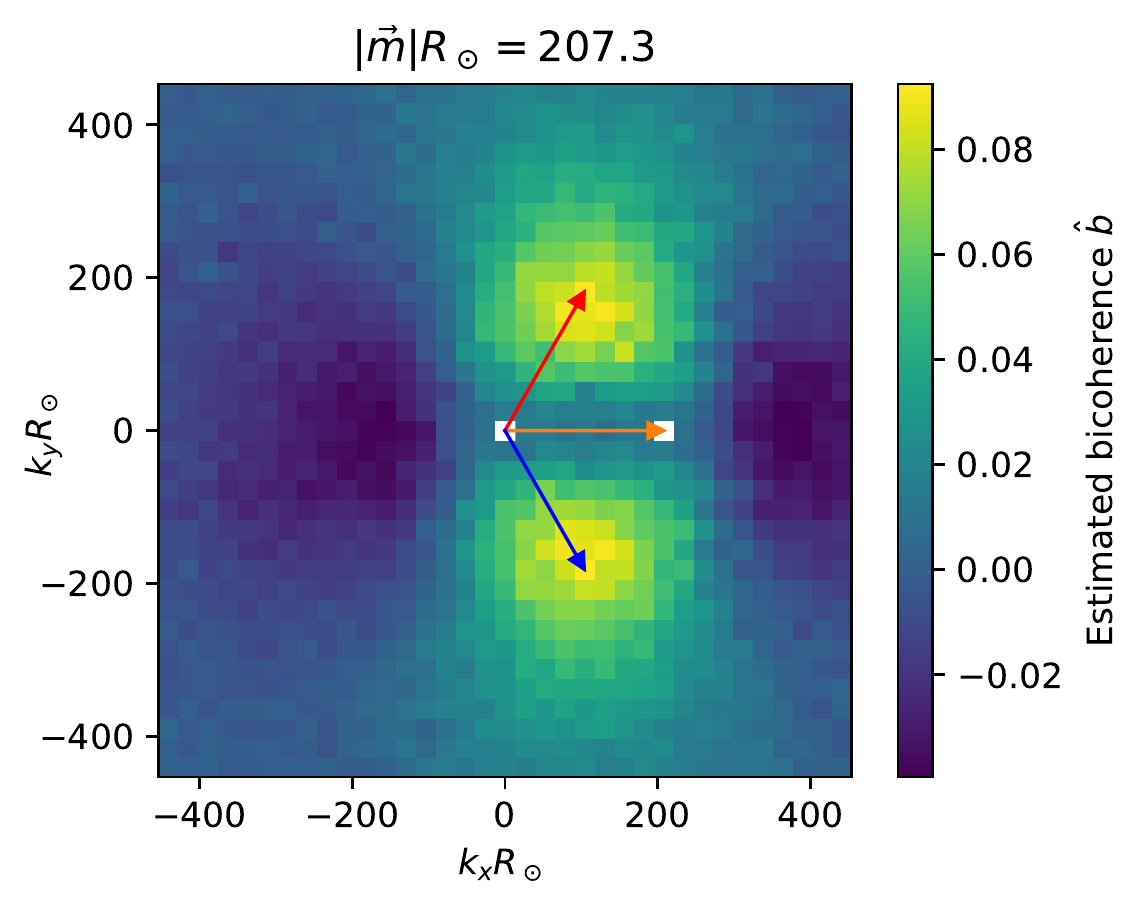}

        \includegraphics[width=0.32\linewidth]{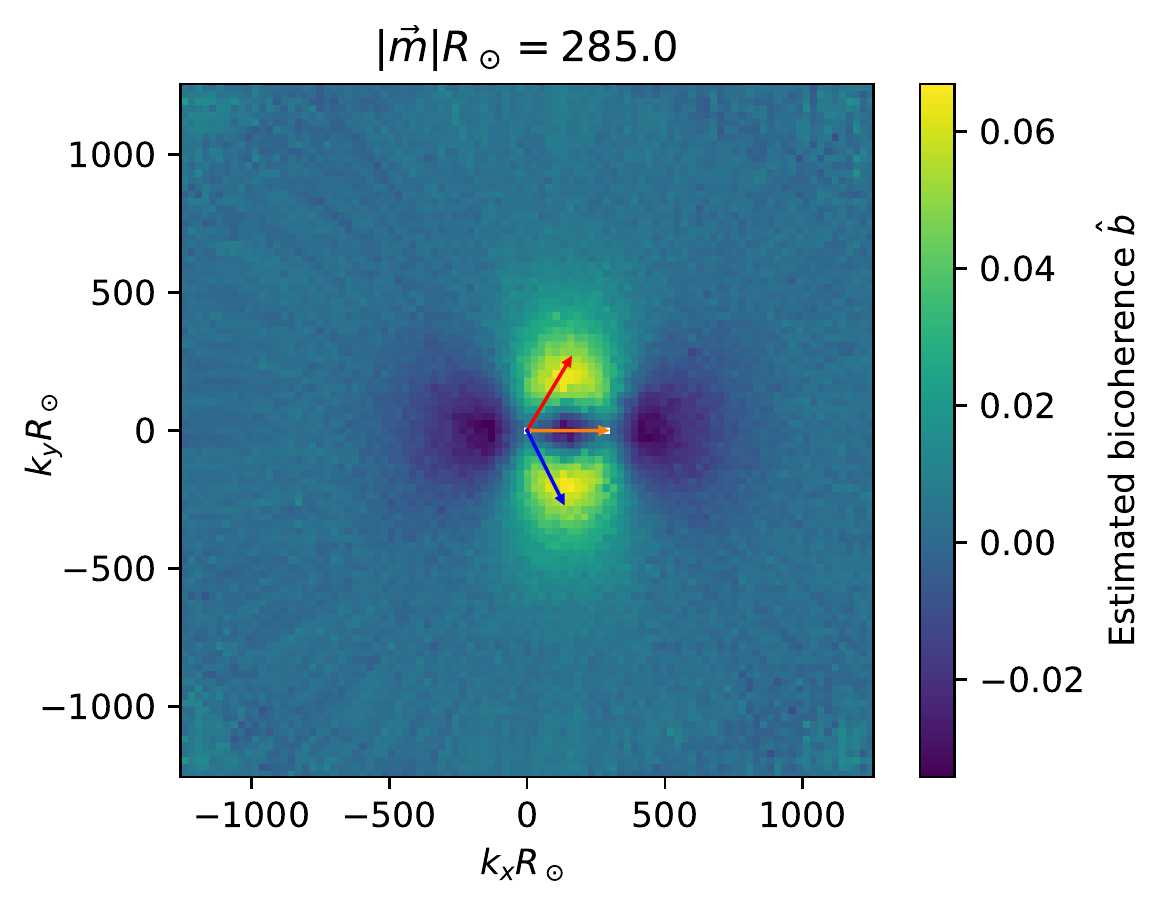}
        \includegraphics[width=0.32\linewidth]{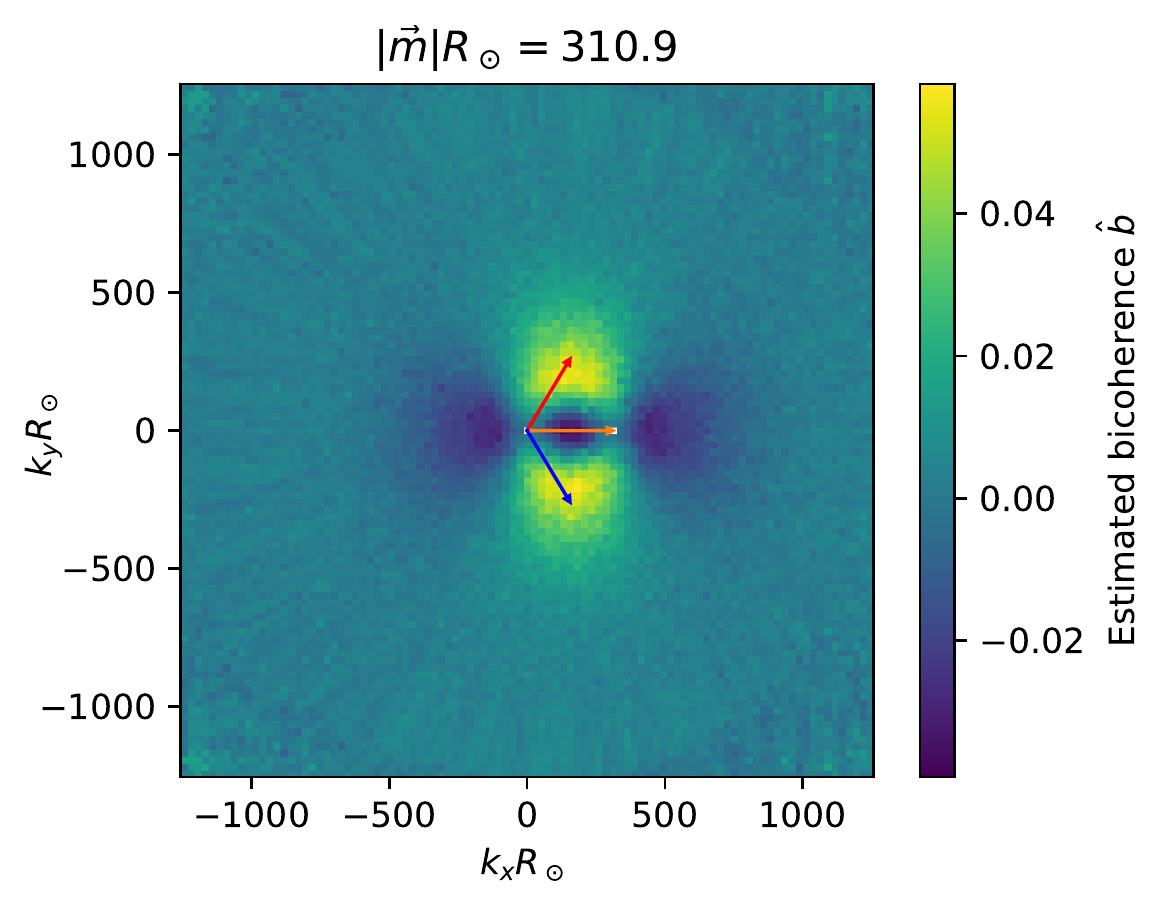}
        \includegraphics[width=0.32\linewidth]{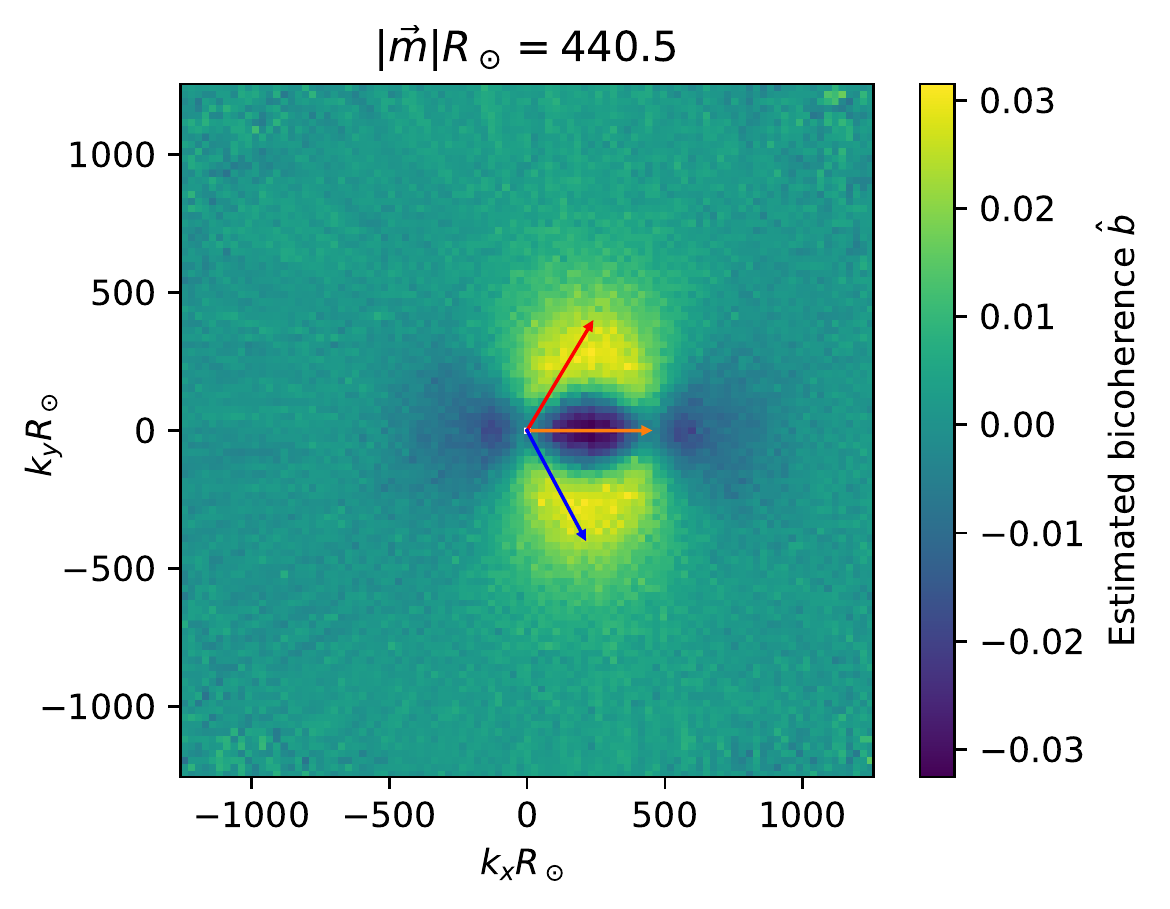}

        \includegraphics[width=0.32\linewidth]{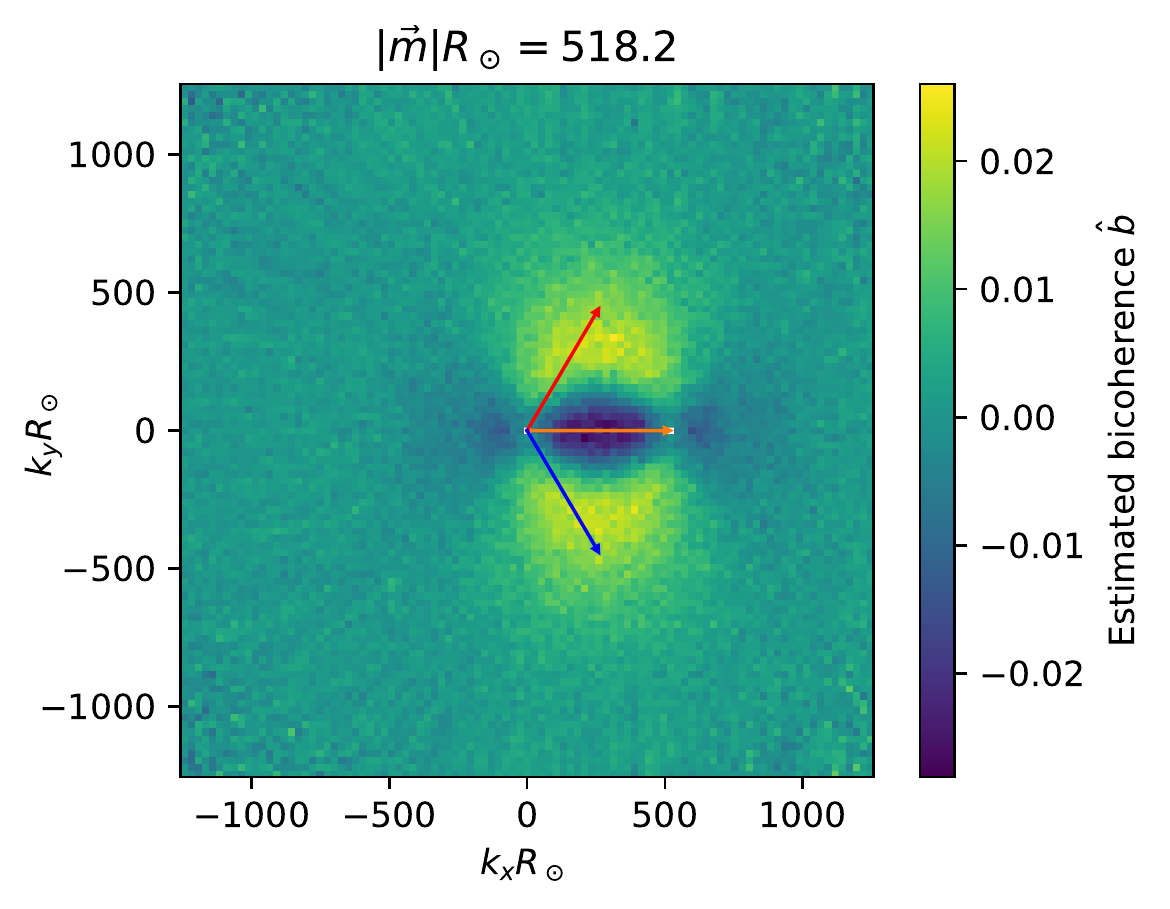}
        \includegraphics[width=0.32\linewidth]{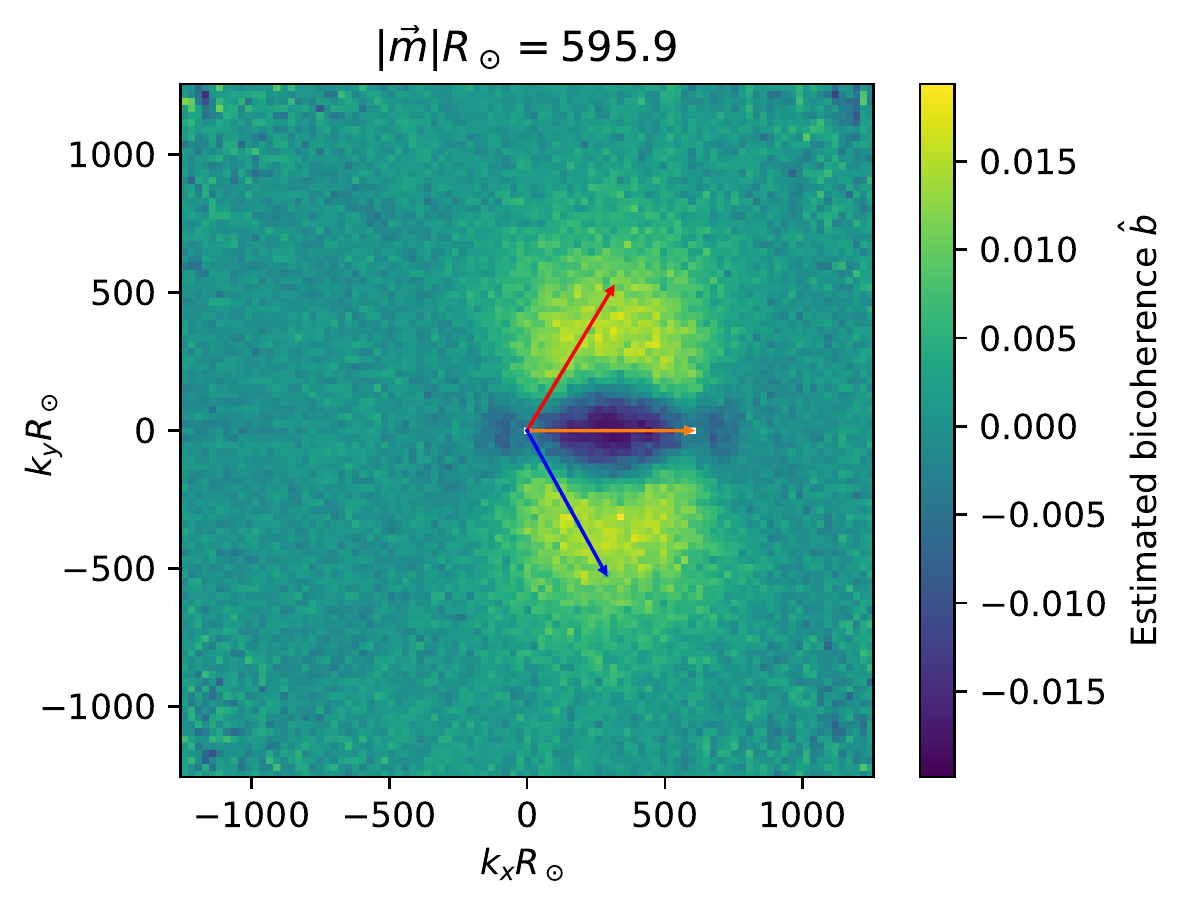}
        \includegraphics[width=0.32\linewidth]{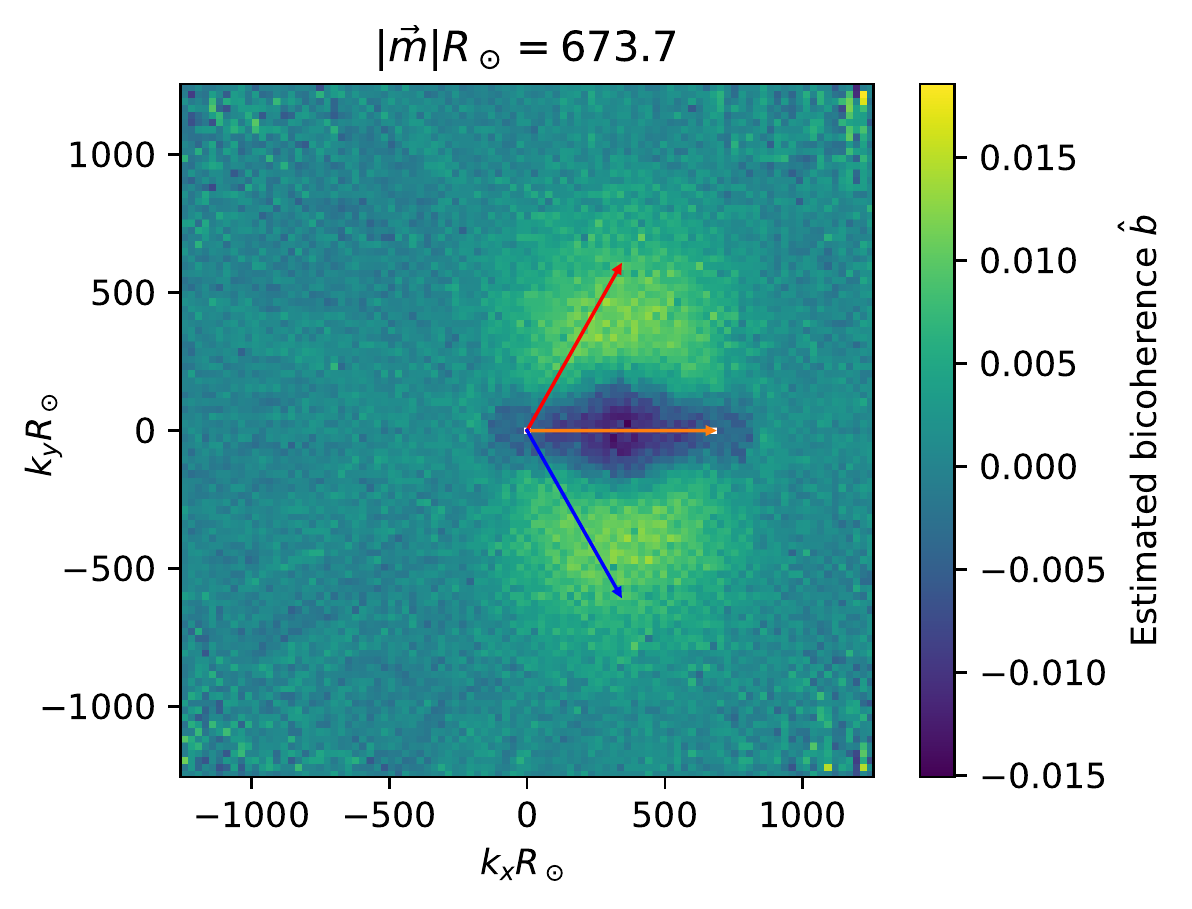}

        \caption{Same as Figure~\ref{figbicoh} for more values of $|\bmm| R_\sun$. We note that the extent of the axis in the top row is different to the other panels.}
        \label{figbicohall}
\end{figure*}

\begin{figure*}
        \centering
        \includegraphics[width=0.32\linewidth]{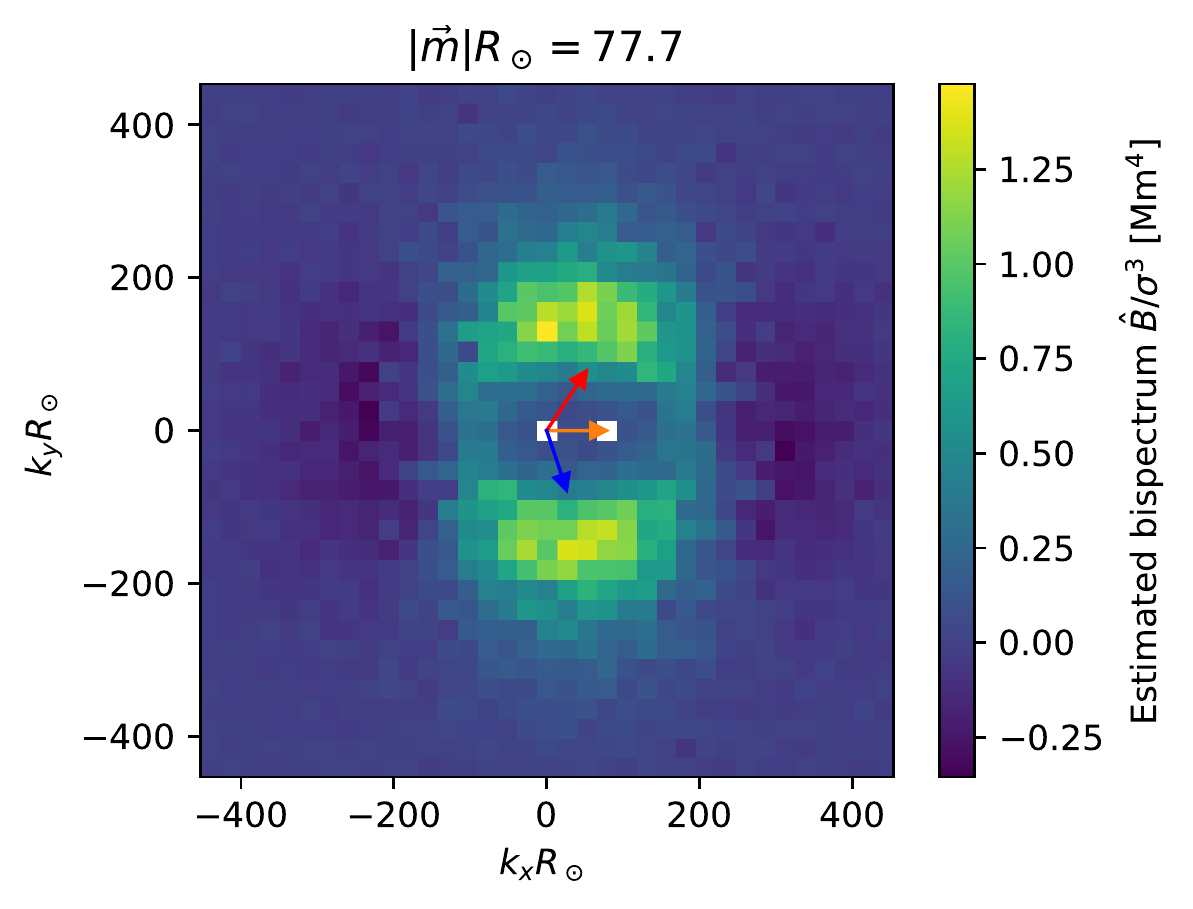}
        \includegraphics[width=0.32\linewidth]{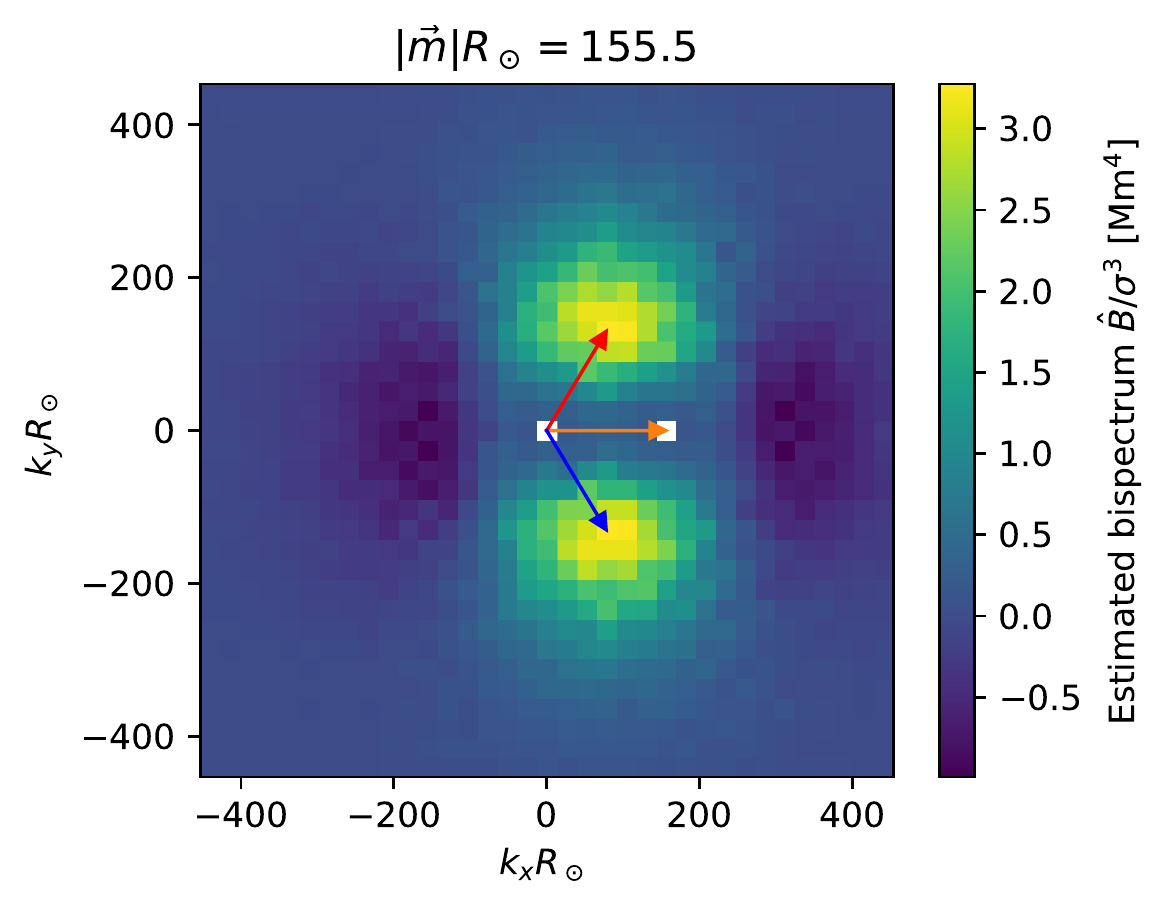}
        \includegraphics[width=0.32\linewidth]{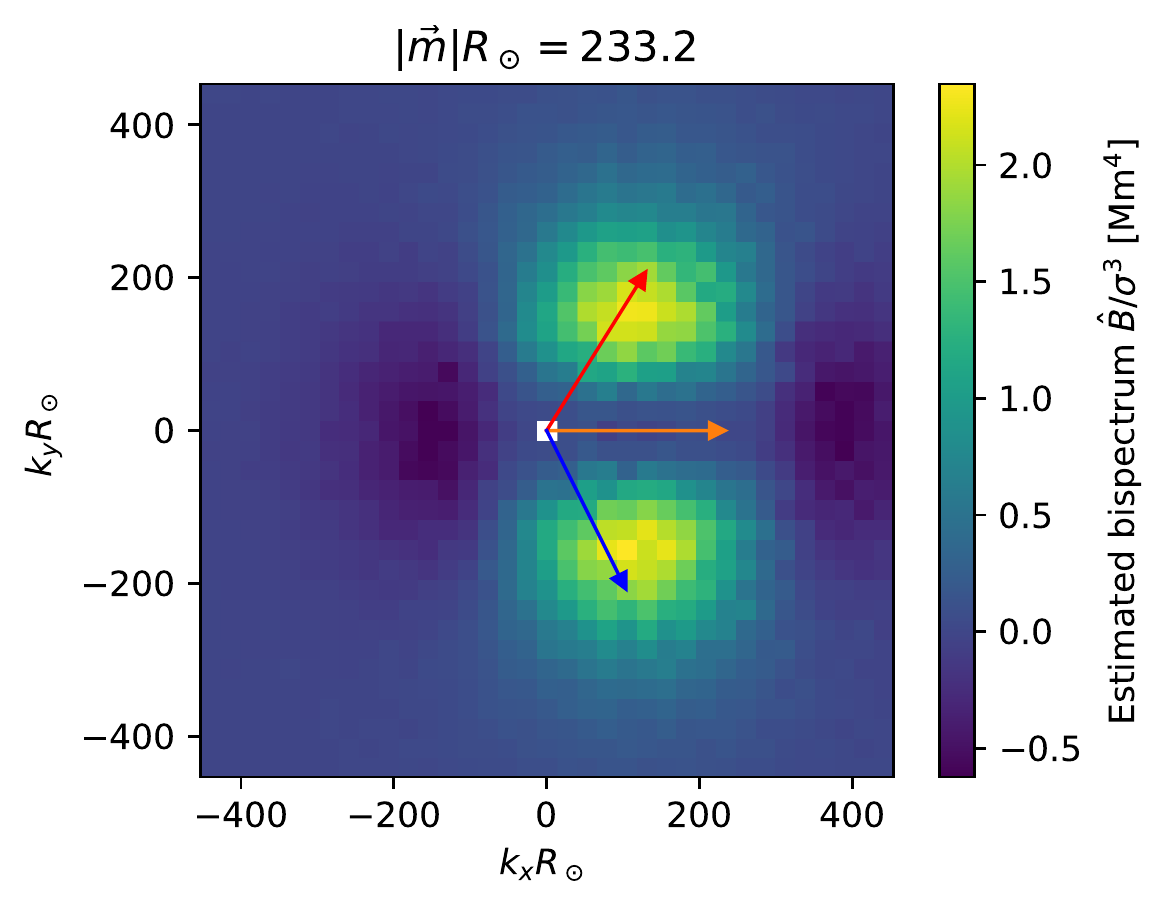}

        \caption{Same as Figure~\ref{figbicoh}, but for the estimated bispectral density $\hat B / \sigma^3$.}
        \label{figbispec}
\end{figure*}

\begin{figure*}
        \centering
        \includegraphics[width=0.32\linewidth]{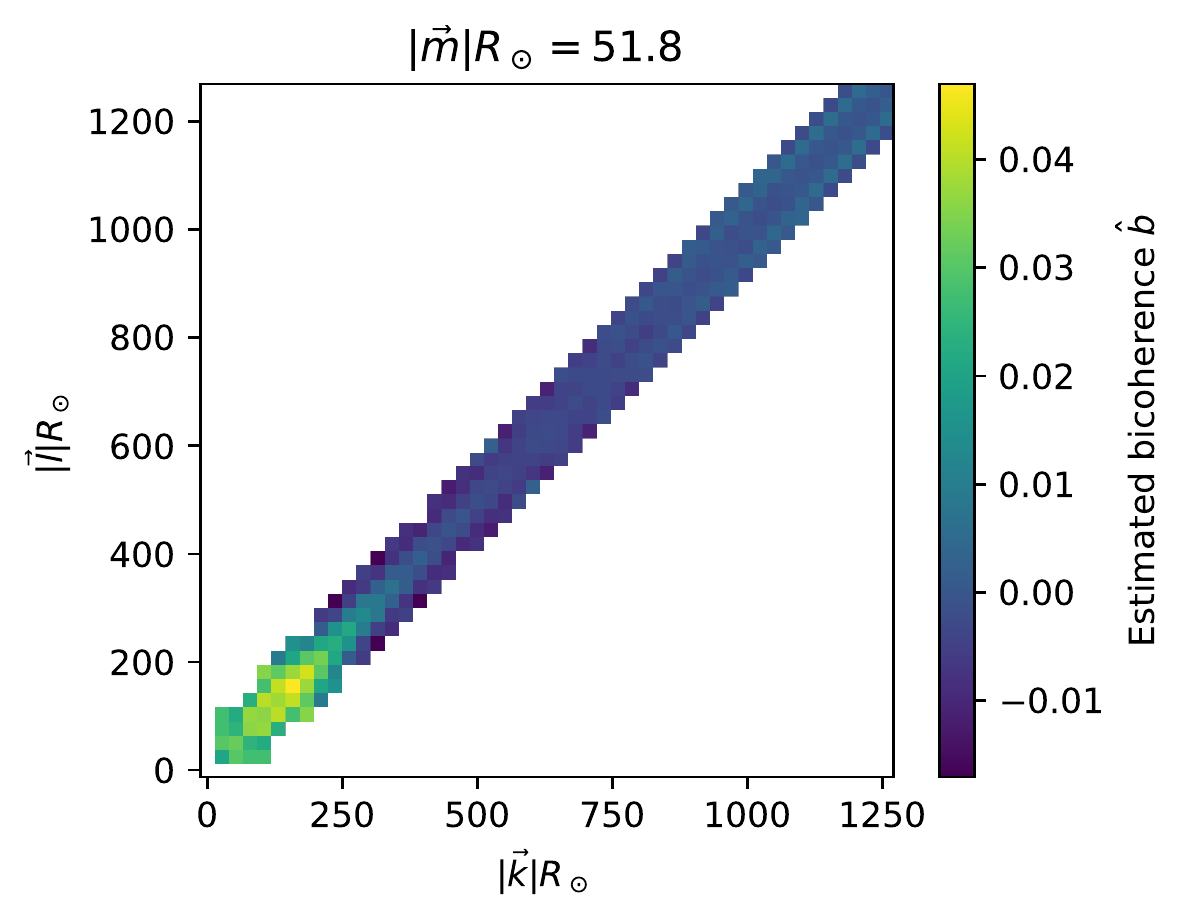}
        \includegraphics[width=0.32\linewidth]{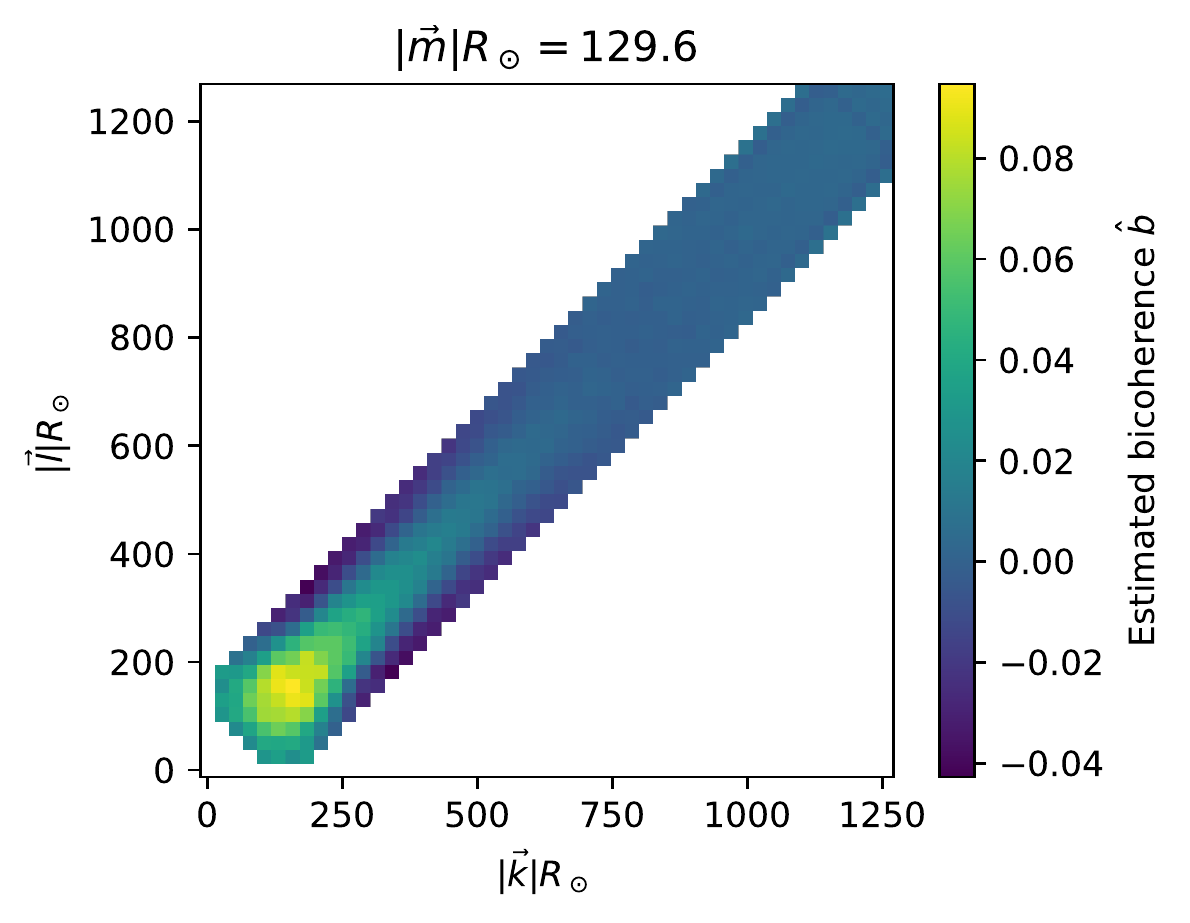}
        \includegraphics[width=0.32\linewidth]{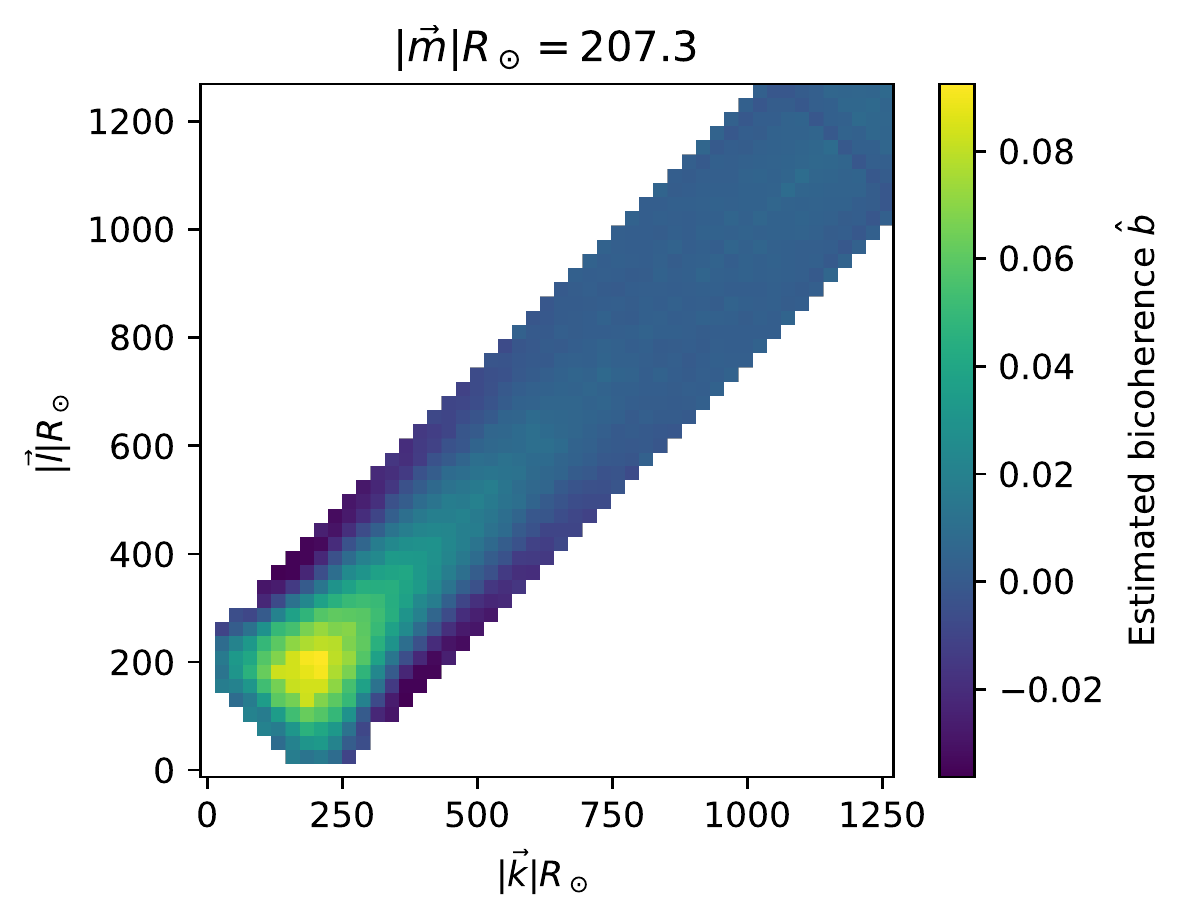}
        
        \includegraphics[width=0.32\linewidth]{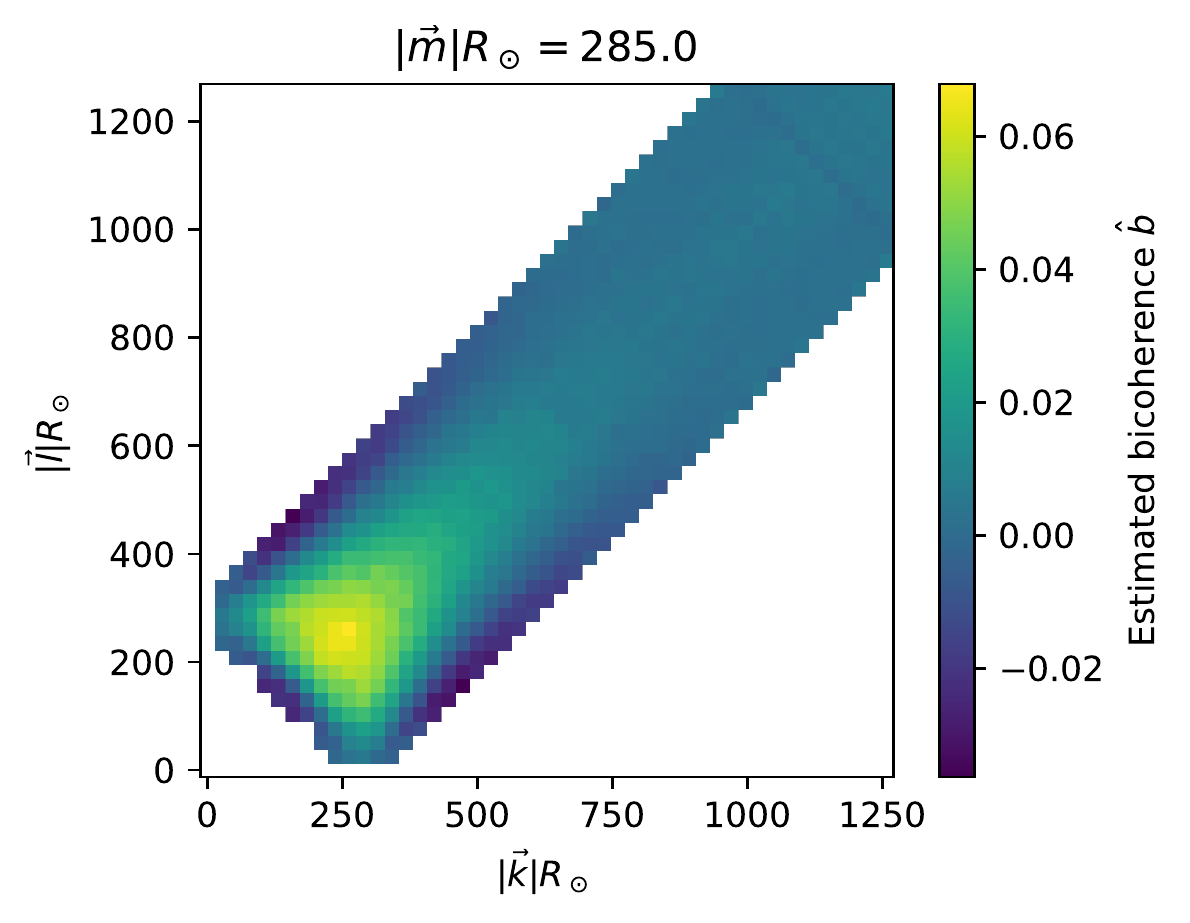}
        \includegraphics[width=0.32\linewidth]{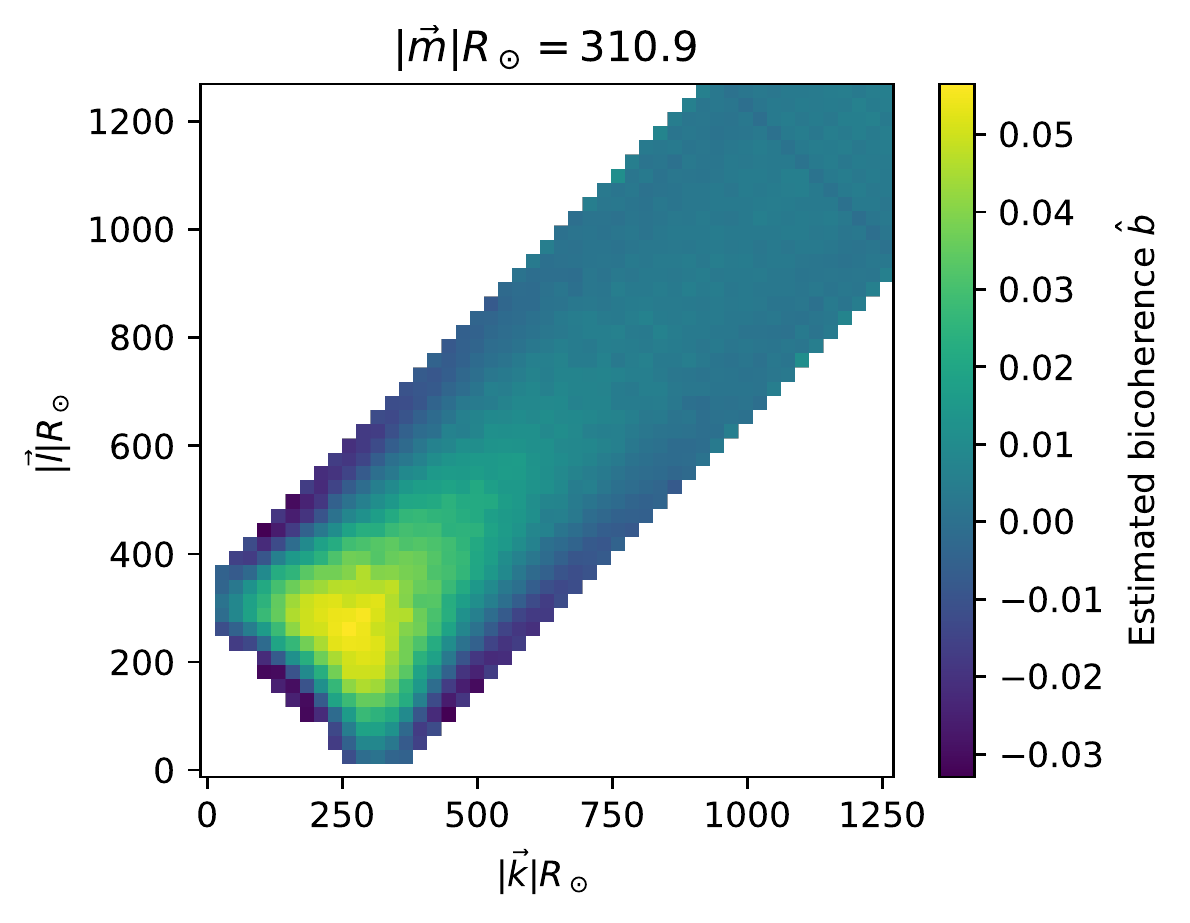}
        \includegraphics[width=0.32\linewidth]{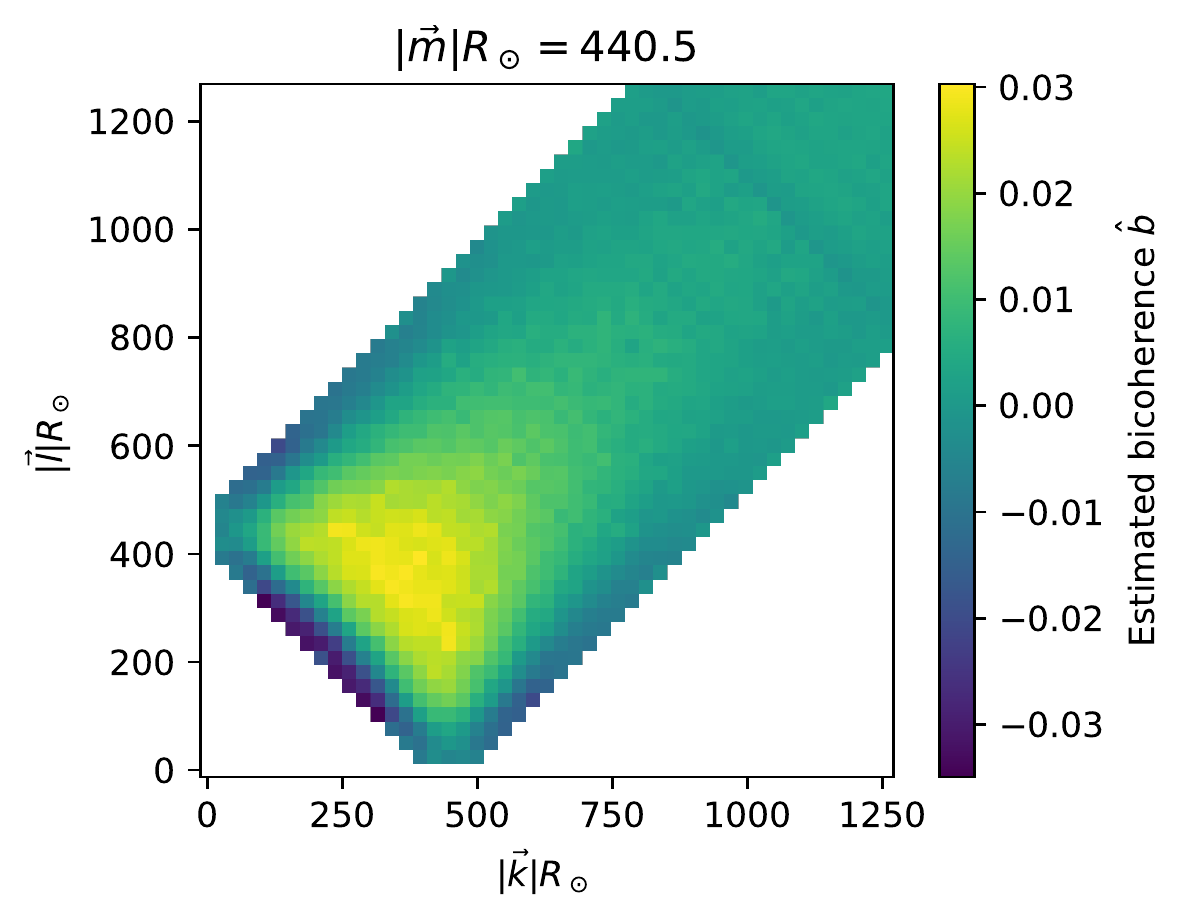}
        
        \includegraphics[width=0.32\linewidth]{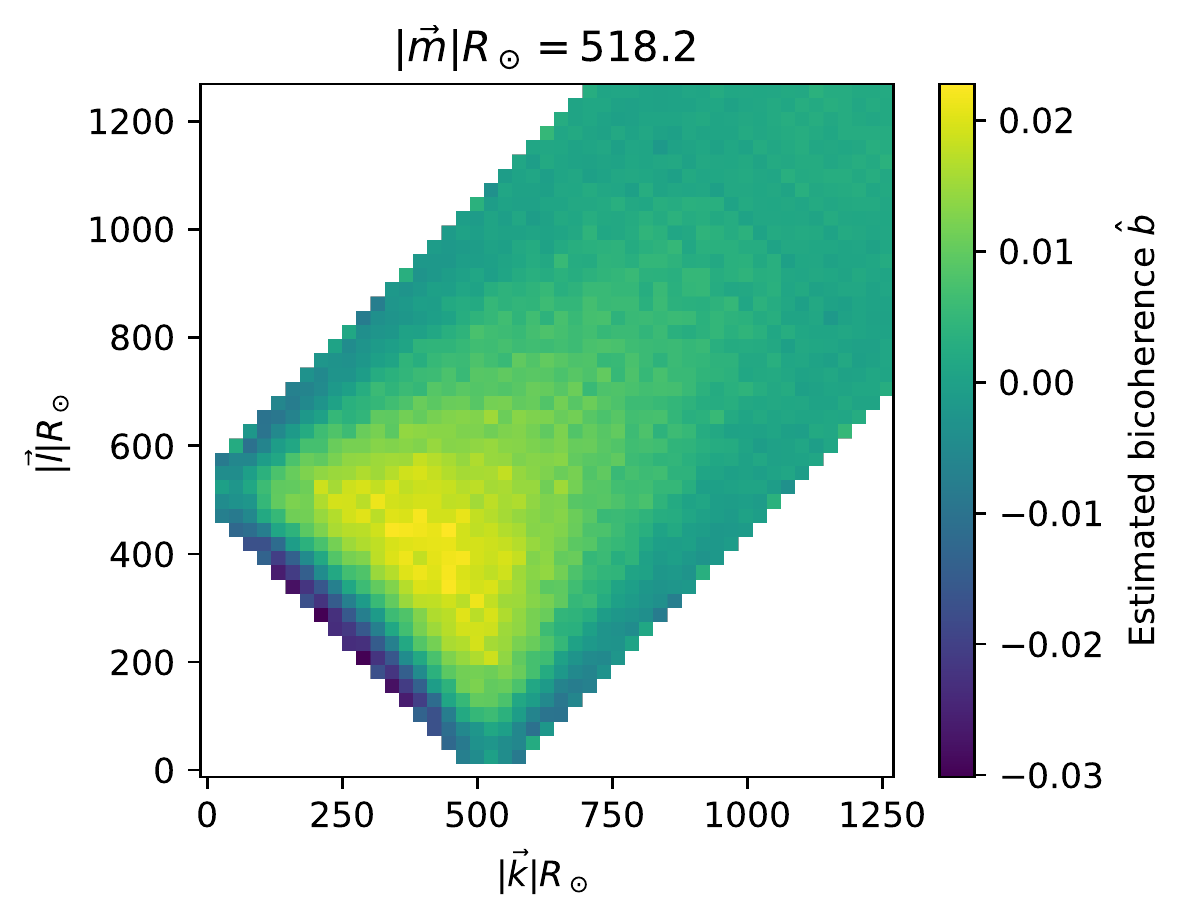}
        \includegraphics[width=0.32\linewidth]{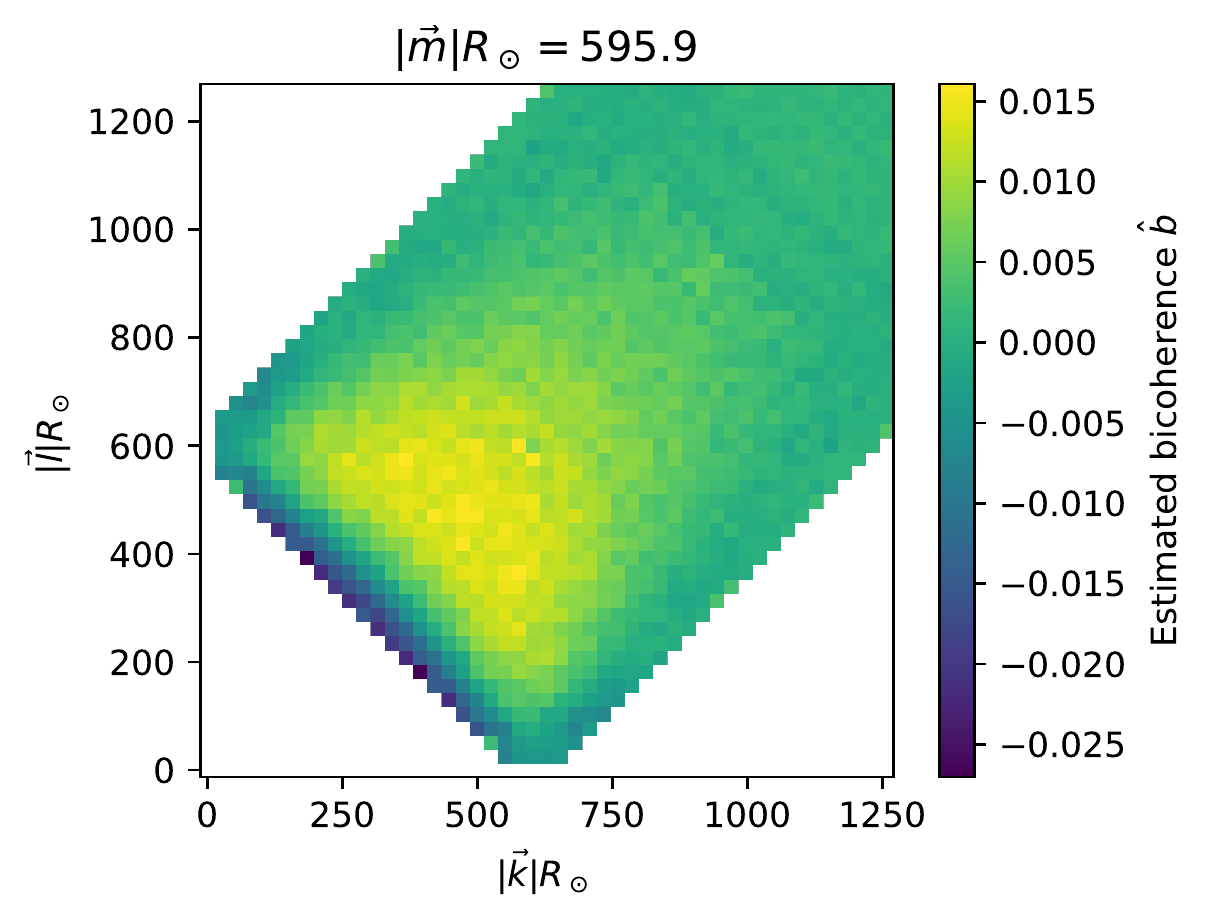}
        \includegraphics[width=0.32\linewidth]{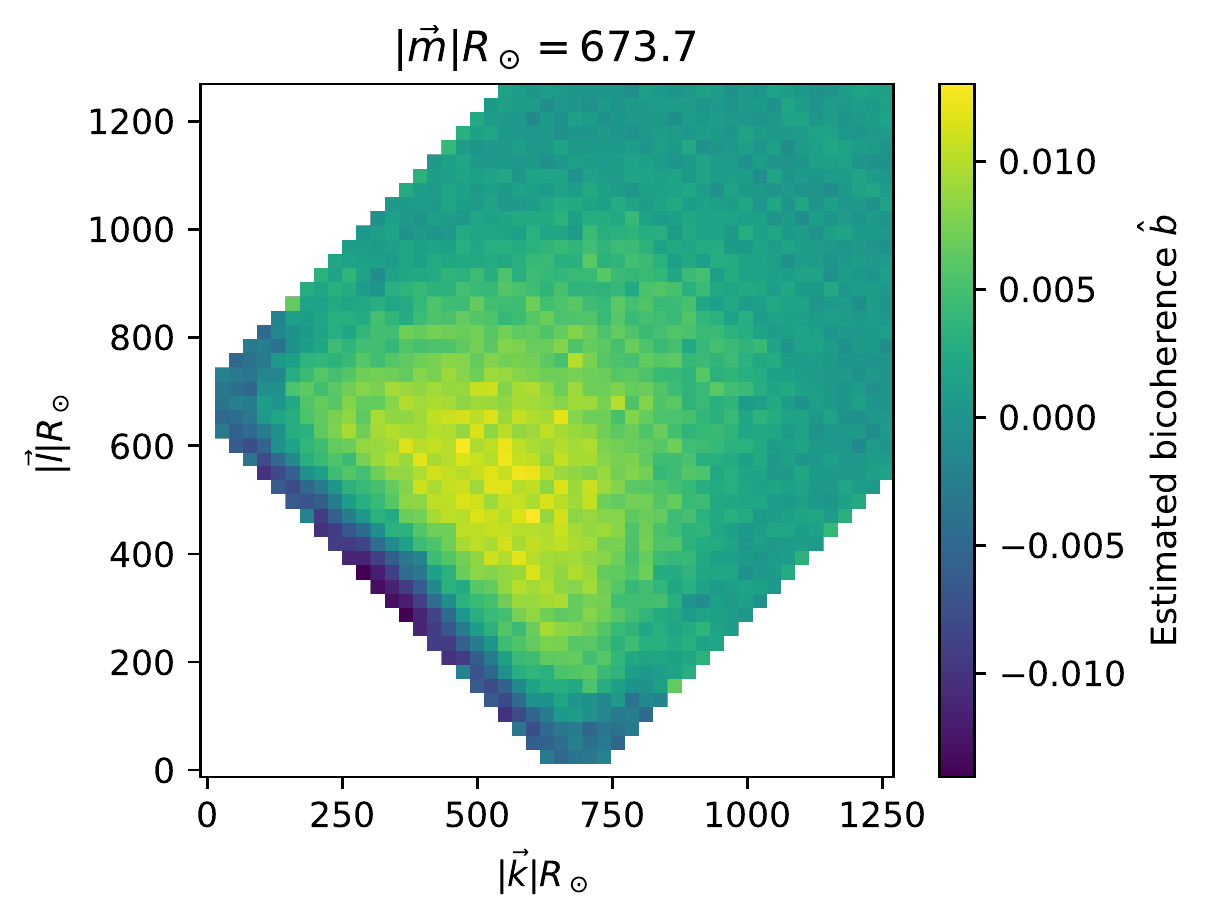}

        \caption{Estimates of the bicoherence $b$ from LCT as a function of $(|\bk|,|\bl|)$ for the same values of $|\bmm|$ as in Figure~\ref{figbicohall}, also noted in the top of each panel. White color denotes locations where either the triadic relation $\bk+\bl=\bmm$ cannot be met or where one of the three vectors has zero length (as the mean was removed from each map, the bicoherence is not defined in this case).}
        \label{figbicohKLall}
\end{figure*}

\begin{figure*}
        \centering

        \includegraphics[width=1\linewidth]{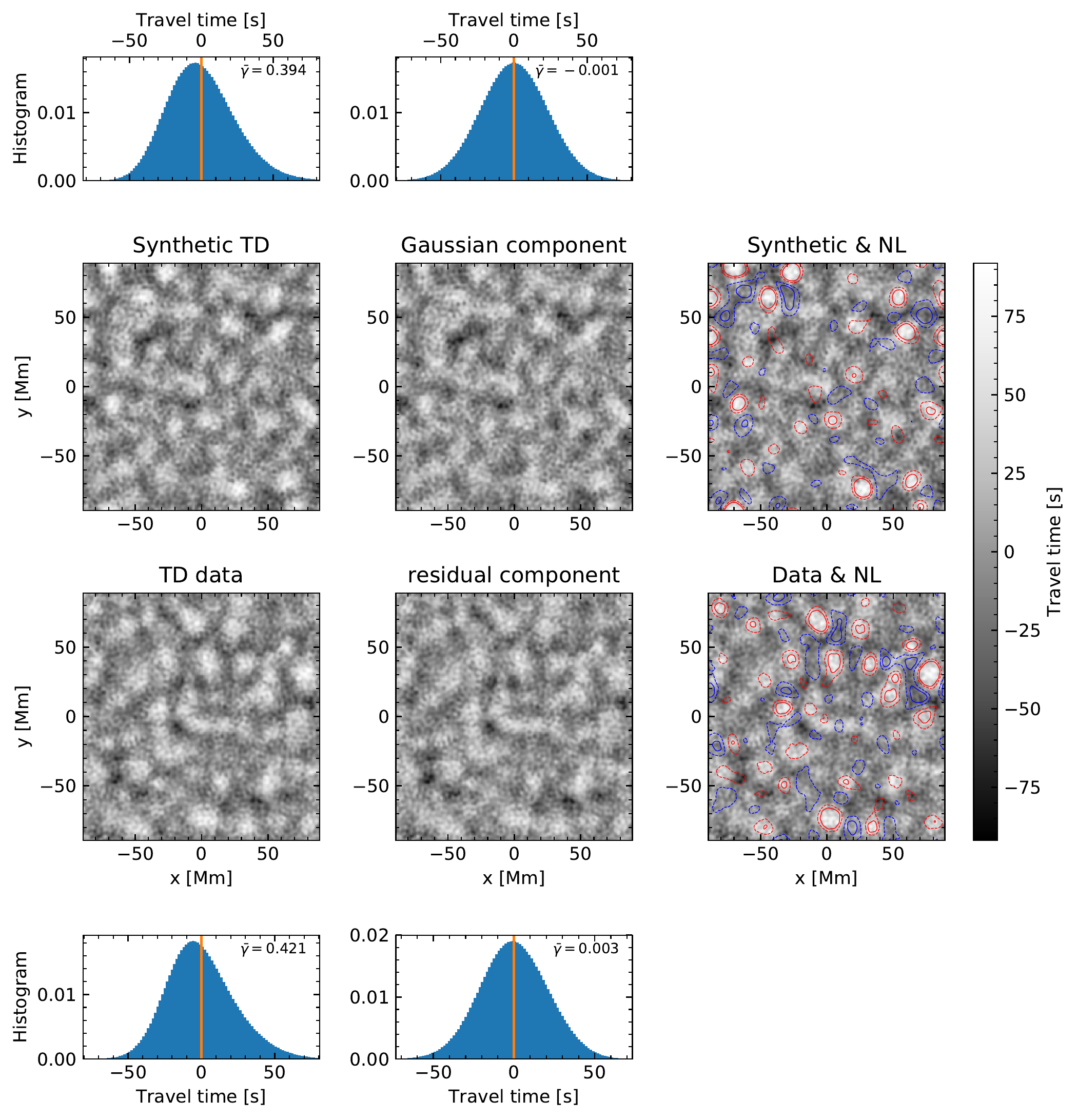}

        \caption{Same as Figure~\ref{figsynthexample}, but for TD data.}
        \label{figsynthexampleTD}
\end{figure*}

In Figure~\ref{figbicohall}, we show results for the bicoherence for more values of $|\bmm| R_\sun$ as shown in Figure~\ref{figbicoh}. Figure~\ref{figbispec} shows results for the bispectrum. Finally, in Figure~\ref{figbicohKLall}, we show results for the bicoherence as a function of the amplitude of all coupling Fourier vectors, that is of $(|\bk|,|\bl|,|\bmm|)$, for the same selection of $|\bmm|$ as in Figure~\ref{figbicohKLall}.

\end{document}